\documentclass[lettersize,journal]{IEEEtran}
\usepackage{amsmath,amsfonts}
\usepackage{algorithmic}
\usepackage{algorithm}
\usepackage{array}
\usepackage{textcomp}
\usepackage{subfigure}
\usepackage{tabularx}
\usepackage{arydshln}
\usepackage{marvosym}
\usepackage{stfloats}
\usepackage{url}
\usepackage{verbatim}
\usepackage{graphicx}
\usepackage{multirow} 
\usepackage{cite}

\begin{document}

\title{A Knowledge-Driven Cross-view Contrastive Learning for EEG Representation}

\author{Weining~Weng,
        Yang~Gu,
        Qihui~Zhang,
        Yingying~Huang,
        Chunyan~Miao,
        and~Yiqiang~Chen\textsuperscript{*}% <-this % stops a space

\thanks{Weining Weng, Yang Gu and Yiqiang Chen are with the Institute of Computing Technology, Chinese Academy of Sciences, Beijing, China, and Beijing
Key Laboratory of Mobile Computing and Pervasive Device, Institute
of Computing Technology, Chinese Academy of Sciences and University of Chinese Academy of Sciences, Beijing, China. E-mail:  \{wengweining21b, guyang, yqchen\}@ict.ac.cn.}% <-this % stops a space
\thanks{Chunyan Miao is with the with the Joint NTU-UBC Research Centre of Excellence in Active Living for the Elderly (LILY), Nanyang Technological University, Singapore. E-mail: ascymiao@ntu.edu.sg.}
\thanks{Qihui Zhang is with the Department of Neurology, Dong Fang Hospital, Beijing University of Chinese Medicine, Beijing, China. E-mail: qihuizhang@vip.sina.com.}
\thanks{Corresponding author: Yiqiang Chen.}
}

\markboth{Submission to TKDE}%
{Shell \MakeLowercase{\textit{et al.}}: A Sample Article Using IEEEtran.cls for IEEE Journals}

\IEEEpubid{}

\maketitle

\begin{abstract}
Due to the abundant neurophysiological information in the electroencephalogram (EEG) signal, EEG signals integrated with deep learning methods have gained substantial traction across numerous real-world tasks. However, the development of supervised learning methods based on EEG signals has been hindered by the high cost and significant label discrepancies to manually label large-scale EEG datasets. Self-supervised frameworks are adopted in vision and language fields to solve this issue, but the lack of EEG-specific theoretical foundations hampers their applicability across various tasks. To solve these challenges, this paper proposes a knowledge-driven cross-view contrastive learning framework (\textbf{KDC2}), which integrates neurological theory to extract effective representations from EEG with limited labels. The KDC2 method creates scalp and neural views of EEG signals, simulating the internal and external representation of brain activity. Sequentially, inter-view and cross-view contrastive learning pipelines in combination with various augmentation methods are applied to capture neural features from different views. By modeling prior neural knowledge based on homologous neural information consistency theory, the proposed method extracts invariant and complementary neural knowledge to generate combined representations. Experimental results on different downstream tasks demonstrate that our method outperforms state-of-the-art methods, highlighting the superior generalization of neural knowledge-supported EEG representations across various brain tasks.
\end{abstract}


\section{Introduction}
\IEEEPARstart{E}lectroencephalogram (EEG) signal is the collected physiological signal generated directly from the human neural activity \cite{r1}. Neuroscience research shows that brain activities induced by different events cause the vertebral neurons to generate a potential at the post-synapse. The potential with the same direction generated by a large number of pyramidal neurons is transmitted to the surface of the cerebral cortex through capacitive conduction and volume conduction, which can be captured by an external electrode on a certain scalp in the form of a weak voltage \cite{r2}. The multi-channel EEG signal measures the activity of billions of neurons inside the brain from the spatial perspective and contains abundant neurophysiological information, so it has been widely used to study brain function and dysfunction, including neurological disorders such as epilepsy \cite{r4}, sleep disorders \cite{r5} and traumatic brain injury \cite{r6}. Its non-invasive nature and ability to measure brain activity in real time make it a valuable asset in research and clinical settings. Due to the high application and clinical value of EEG signals, analyzing and mining neural features related to different brain tasks (neurological disorders, cognitive) from EEG signals has practical significance.
\begin{figure}[t]
  \centering
  \includegraphics[width=\linewidth]{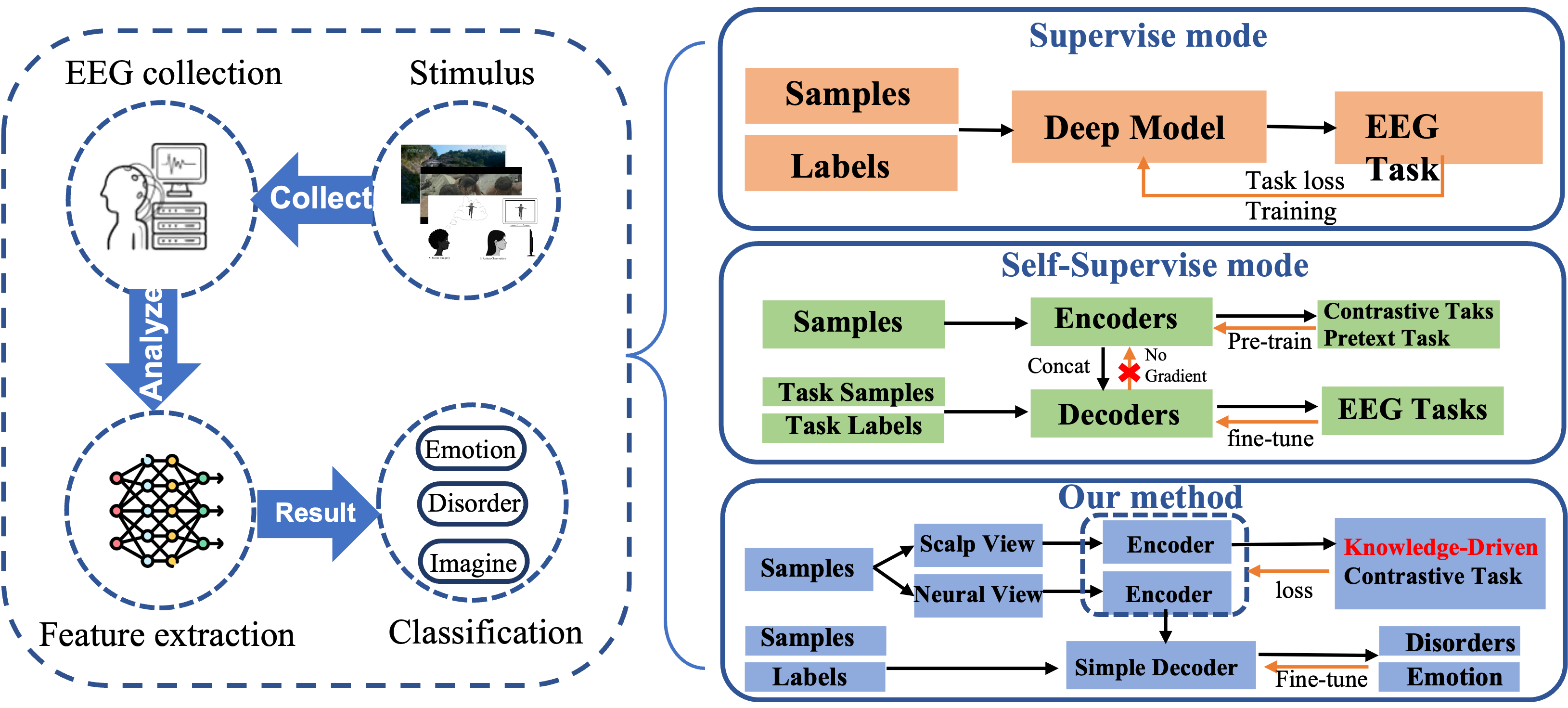}
\caption{Comparison of different EEG analysis methods: the supervised method, the self-supervised method, and our knowledge-driven self-supervised method.}
  \label{figure_1}
\end{figure}

With the development of artificial intelligence, the EEG signal experienced a three-stage transition from the statistical analysis stage into the traditional machine learning and deep learning stage. In the early stage, neuroscientists mainly performed EEG signal analysis manually. They adopted some statistical methods to analyze the significant differences in EEG event-evoked potentials between subjects with different brain states and extract handcrafted neural features correlated with specific diseases and brain states \cite{r8}. In the traditional machine learning stage, researchers combined the handcrafted features with some machine learning models\cite{r9} to automatically solve some simple tasks based on a small amount of labeled training data. Most importantly, the deep learning methods have brought significant advancements to EEG analysis, making it easier and more efficient to use EEG for diagnosis and research\cite{r12}. The end-to-end spatial-temporal deep learning methods \cite{r15} can learn more efficient brain features from EEG signals that are related to different brain functions. Adequate labeled EEG data and powerful deep learning models are the two critical elements.

The most critical challenge in combining the deep learning method into different EEG tasks is the labeling issue \cite{r16}. The supervised deep learning methods require a large amount of labeled data for model training to extract task-related features. In computer vision and natural language processing, the image and text data are readily available with a low manual annotation threshold. Furthermore, the variability in annotation and labeling error rates for image and text data are generally small, which makes these data types more amenable to human annotation and facilitates their use in deep learning and other data-driven applications. However, EEG signals are complex and dynamic, and their interpretation requires specialized knowledge and expertise in neurophysiology, signal processing, and cognitive neuroscience, leading to difficulties and high costs for large-scale labeling of task-specific EEG data. Additionally, there may be variability in EEG signals across individuals and conditions, which makes it challenging to establish consistent labeling criteria. Therefore, it is challenging and meaningful to mitigate the effects of the labeling difficulties and labeling noise and extract task-related features from EEG data for different less-labeled EEG tasks.

Recent researchers have combined representation learning methods with EEG signals to generate effective representations from EEG samples with the aid of limited labels \cite{r17}. Different pretext tasks and contrastive tasks can help the model to learn abstract features from the EEG signal: the mask-reconstruct task\cite{r18}, the spatial-temporal jigsaw task \cite{r19}, and the sample augmentation contrastive task \cite{r20}, etc. The representation learned by the pretext and contrastive self-supervised tasks mentioned above can help the model improve its effectiveness on specific downstream tasks. However, those methods are mere extensions of self-supervised frameworks in computer vision and language tasks. These self-supervised tasks explore the spatial and contextual semantics of multi-channel EEG signals according to image or text data patterns while ignoring the EEG signals' inherent neural mechanism and complex dynamic neural correlations. Due to the lack of theoretical support for prior neural knowledge, existing EEG representation methods face difficulties in extracting generalized neural features that can be applied to different downstream tasks. Therefore, integrating the neural prior knowledge of EEG signals in the model and mine for more generalized representations remains a key challenge.

The above challenges demonstrate that the traditional supervised learning models and the existing self-supervised models are unsuitable for extracting neural representations because they rely too much on the labeled samples or ignore the neural features of the EEG signal. Hence, it is of vital importance to design an innovative self-supervised framework that combines the EEG neural theory to extract general neural representation applicable to various EEG tasks, which needs to carefully design the pretext or contrastive task to mine for neural knowledge and high-level features according to the neural theory from EEG signals. Critical issues need to be addressed in this new SSL (Self Supervised Learning) framework: (1) How to learn more general and efficient features from EEG data with limited labels instead of the massive labeled samples. (2) How to combine the neural theory in the pretext or contrastive task to extract the explainable knowledge-level features instead of the data-level features.    
 
Inspired by the multi-view learning widely used in computer vision \cite{r21}, graph classification \cite{r22}, and recommendation system \cite{r24}. This paper innovatively proposes a knowledge-driven cross-view contrastive learning method, a new framework to learn neural representation from EEG signals applicable to different brain tasks. Firstly, the model constructs two different views for EEG signal according to the neural theory of EEG generation: the cerebral cortex view describes the spatial distribution of voltage collected by external electrodes on the brain scalp, and the neural topological view describes the topological correlation of intrinsic neural dipoles in EEG signals. Multiple data augmentation methods are applied to expand samples in each view for contrastive, where the augmentation methods used in the two views are the same. Spatial-oriented convolutional neural network encoders and topology-oriented graph convolutional encoders mine EEG representations from corresponding views. The non-negative contrastive loss is well-designed inside different views to help extract the invariant features from variation augmented samples. Inter the view, according to the EEG generation theory that different views are the internal and external representations of neural activity which are essentially neurally homologous, the main idea of the cross-view contrastive methods can be simplified as "contrast with same augmentation" to model the prior neural knowledge: samples with same augmentation methods from different views are considered to have highly similar neural information and form the positive pairs, while samples with different augmentation methods from different views are significantly discrepant and form the negative pairs. The cross-view contrastive method extracts neural knowledge from similar information expressed by the internal and external view of brain activity. The minimization of the comprehensive loss function that combines multiple contrastive losses can self-supervised train the model and generate the fused EEG neural representation combined with the complementary neural knowledge from each view. The further experimental results demonstrated that the proposed method outperformed SOTA models in various downstream EEG tasks on both pre-train and joint-train modes, which proves the efficiency and generalization of the learned neural representations. The contributions of this paper are summarized as follows:

 (1) This paper  proposes the cross-view contrastive learning framework to extract neural representations from EEG signals. The framework learns the neural activity features through the collected samples with few labels, which have important practical significance in various real EEG application scenarios with limited labels.

 (2) The proposed method incorporates the neural theory of EEG signal into the contrastive task. The combined contrastive learning methods model prior neural knowledge and extract invariant and dynamic neural features from different views according to homologous neural information consistency theory. The knowledge-driven contrastive framework is the first method to understand neural semantics and information, with stronger expressive and generalization capabilities.
 
 (3) We evaluated the framework on three different EEG tasks. The experimental results indicate that our method outperforms SOTA self-supervised learning and traditional supervised learning algorithms in all tasks on the pre-train and joint-train modes, which verifies the superior performance of the proposed framework.

\section{Related Works}
\subsection{EEG-based brain tasks}
The EEG signal contains massive neural information and brain activity features with important clinical values. Many researchers recently combined adequately collected EEG samples with powerful deep models to address different EEG-based brain tasks. 

Many researchers focused on using EEG signals for emotion recognition and extracting the emotion-related EEG features: Kshare et al. proposed to use a deep convolutional neural network to extract emotion-related components from the time-frequency domain of EEG features, which can help to accurately identify four types of emotion from EEG signal \cite{r26}; Tao et al. constructed a deep end-to-end model that combined the CNN, LSTM, and channel-wise attention mechanism to recognize discrete and continuous emotions and adaptively select critical emotion-related channels \cite{r27}.

Besides, the motor imagery (MI) task is another important EEG-based task that has been widely investigated by researchers. The MI task can be defined as the neural states generated by the subject mentally imagining the physical action \cite{r29}. Amin et al. \cite{r30} proposed to use the cascade deep convolutional neural network to extract the neural features from the EEG row data to identify the imagined movements of different limbs (right hand, left hand). Hou et al. \cite{r32} exploited the topological correlation of electrodes and proposed a deep framework that combined the graph convolution neural network and graph pooling to improve the decoding performance for different EEG-based MI tasks. 

Most importantly, EEG signals have a very wide range of clinical tasks. Different brain-related disorders are investigated and studied through EEG signals: Srirangan et al. \cite{r35} carried out research on the time-frequency domain of EEG signals, they combined the handcrafted features extracted by fast Fourier transform and the wavelet transform with CNN to locate the epileptogenic regions during epilepsy; Sun et al. \cite{r36} proposed a hybrid deep neural network adopts the CNN-RNN structure to classify schizophrenia from frequency-domain features of EEG signal. Besides, different brain-related disorders like mild cognitive impairment, Alzheimer's disease, and Parkinson's have also been widely studied by researchers. In summary, the EEG signals combined with powerful deep learning methods such as the CNN, RNN, graph neural network, and Transformer have shown good performance in different EEG-related tasks, promoting automatic EEG-based brain task application.

\begin{figure*}[ht]
  \centering
  \includegraphics[width=\textwidth]{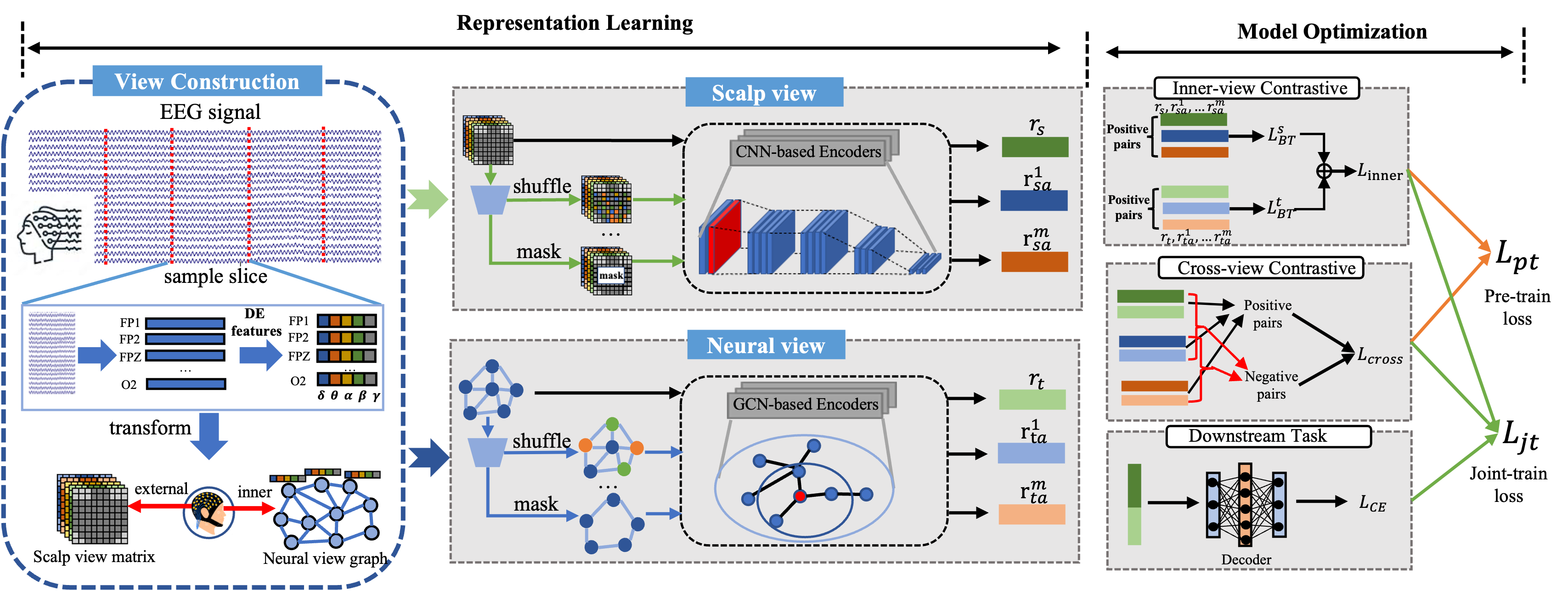}

  \caption{The detailed structure of the proposed KDC2 method. The proposed method contains three parts: 1. The view construction part, which is used to preprocess the collected EEG signal, extract sample preliminary features, and transform into two different views. 2. The representation part, which is used to generate representations from two view separately through different base encoders. 3. The optimization part, which combines the inner-view and cross-view contrastive losses to self-supervised train the model and fused the task-specific loss to joint train the model.}
  \label{fig2}
\end{figure*}

\subsection{Self-supervised learning in EEG}
With the success of self-supervised learning frameworks in traditional fields such as CV (Computer Vision) and NLP (Natural Language Process), many researchers have adopted the idea of self-supervised learning in EEG analysis to learn efficient representations from few-labeled EEG data. Three different self-supervised methods have been extensively investigated to mine representations from multi-channel EEG signals: 1.The discriminative pretext task; 2. The generative pretext task; 3. The contrastive task. The technique details of different SSL methods are listed as follows:

The discriminative pretext tasks set classification or recognition tasks that derive labels from the data. Banville et al. \cite{r37} designed two different discriminative pretext tasks to learn the representations: The relative positioning task to determine whether the sampled EEG fragments are adjacent in the temporal dimension. The temporal shuffle task to shuffle the EEG clips and determine whether the EEG samples remain in the original temporal order. Xu et al. \cite{r39} performed various signal transformations on the EEG data and designed the pretext task of predicting the transformation method to pre-train the model. Li et al. \cite{r19} combined the spatial and frequency jigsaw tasks to jointly pre-train the graph model and extract the EEG representations that incorporate spatial and temporal features, greatly improving the EEG emotion recognition downstream task performance. 

The generative pretext tasks are most widely used in the language model, which adopted the mask-reconstruction or the sequential prediction to help the model understand the contextual correlations with the self-constructed annotations. Multiple generative pretext tasks have been designed to work with EEG signals: Kostas et al. \cite{r41} imitated the well-known language model BERT and conducted the mask-autoencoder structure to extract representations: the transformer encoder inputs the locally masked EEG signals to extract features and representations, while the transformer decoder reconstructs the masked signals with the extracted features. The authors pre-trained the model on the large-scale unlabeled TUH EEG dataset, which can efficiently improve the performance on different downstream tasks (motor imagery, sleep stage, event recognition). Li et al. \cite{r42} proposed a multi-view mask autoencoder framework to reconstruct the masked content in the spectral, spatial, and temporal perspectives of EEG signals to extract emotion-related EEG representations. 

The contrastive tasks aim to contrast samples against each other to teach the model to find common or different data points in the samples, which can help generate general attributes from data directly. Different contrastive methods have been applied to process EEG signals: Mohsenvand et al. \cite{r43} promoted the idea of Simclr and conducted different data augmentation methods (mask, time shift, and band-pass filter.) to construct the positive and negative sample pairs that guide the encoder of CNN-RNN structure to learn more general representation that can improve the model performance on emotion recognition and disorder detection tasks. Kan et al. \cite{r44} designed a meiosis augmentation method that randomly exchanged part of the data between samples as positive pairs, using group-level contrastive learning to extract the representation to recognize emotions. Shen et al.\cite{r45} carried out the cross-subject contrastive learning that sampled negative pairs from different subjects during the same event and positive pairs from the same subject to help the model learn the general EEG representations that minimize the inter-subject differences. All three kind of SSL tasks have successfully improved the training efficiency of EEG signals, but  lacks of the the fitted EEG data for different tasks.

\section{Methdology}
\subsection{Overall Framework}

This paper proposes a knowledge-driven cross-view contrastive learning method (KDC2) that integrates the neural theory of EEG generation to model EEG prior neural knowledge and generate effective neural representations adapted to different downstream brain tasks. Figure \ref{fig2} demonstrates the detailed framework of the proposed method. As illustrated in Figure \ref{fig2}, the method contains five important components: the EEG view construction module, the hybrid data augmentation, the view-independent backbone encoder, the inter-view contrastive learning module, and the cross-view contrastive module. Firstly, the view construction module maps the EEG row signals into the cerebral cortex perspective and neural topological perspective respectively according to the neural theory of EEG generation. Secondly, the hybrid data augmentation method provides consistent augmented samples for different views. Different independent backbone encoders are then designed to extract features from the two views: the convolutional neural network to extract spatial features from the cerebral cortex view and the graph convolutional network to extract topological features from the neural topological view. The inner-view contrastive learning designs a non-negative loss to learn invariant features from positive pairs within views, while cross-view contrastive learning extracts neural knowledge from contrasting the same augmented samples between views. The parameters of the encoders from the two views are jointly optimized by minimizing the multiple contrastive losses. Finally, the neural representations composed of neural features extracted from neural and cerebral cortex views can be adopted to different EEG-based brain tasks through the few-label task fine-tuning.

\subsection{EEG View Construction}
In the first part of the proposed KDC2, this paper aims to construct two views for EEG signals containing homogeneous and complementary neural information. The collected EEG signals are the external voltage representation on the scalp, which is generally used to extract tasks related features. However, the post-synaptic potential generated by the activity of pyramidal neurons in the brain is the source of EEG signals. The potential generates a negative extracellular voltage near the dendrites, which converts the pyramidal neuron into the dipole that can transmit the potential to the surface of the scalp by capacity conduction and volume conduction. Therefore, this paper constructs the internal neural and external scalp views of EEG signals. Firstly, the EEG signals are cut into different time slices $S_i$, where $S_i \in {\Bbb{R}}^{c \times t}$ with $c$ represents the channel number, and $t$ represents the number of sampling point in the slice. Due to former researcher proved that five EEG frequency bands ($\delta$ (0.5-4HZ), $\theta$ (4-8HZ), $\alpha$ (8-12HZ), $\beta$ (12-30HZ) and  $\gamma$ (30-80HZ)) are highly related to the brain neural activity \cite{r46}, this paper adopts the discrete differential entropy \cite{r47} to calculate the preliminary features from five frequency bands of EEG signals, which are calculated as follows:
\begin{equation}
\begin{aligned}
      {DE(S_i)}=\int_{-\infty}^{\infty} \frac{1}{\sqrt{2\pi} \sigma}e^{-\frac{(x-u)^2}{2 \sigma^2}} log(\frac{1}{\sqrt{2\pi }\sigma}e^{-\frac{(x-u)^2}{2 \sigma^2}})\, {\rm d}x \\ =\frac{1}{2}log(2\pi e \sigma^2)
\end{aligned}
\end{equation}
where $\sigma ^ 2$ and $u$ represent the expectation and variance of the input signal. The differential entropy can be simplified into the discrete form by assuming the input temporal signal obeys the Gauss distribution $N(u,\sigma^2)$. The DE features of five frequency bands are calculated for each channel in the EEG signal, forming the preliminary features matrix $M \in {\Bbb{R}}^{c \times 5}$. Based on the preliminary features, two different views are then constructed.
\begin{figure}[t]
  \centering
  \includegraphics[width=\linewidth]{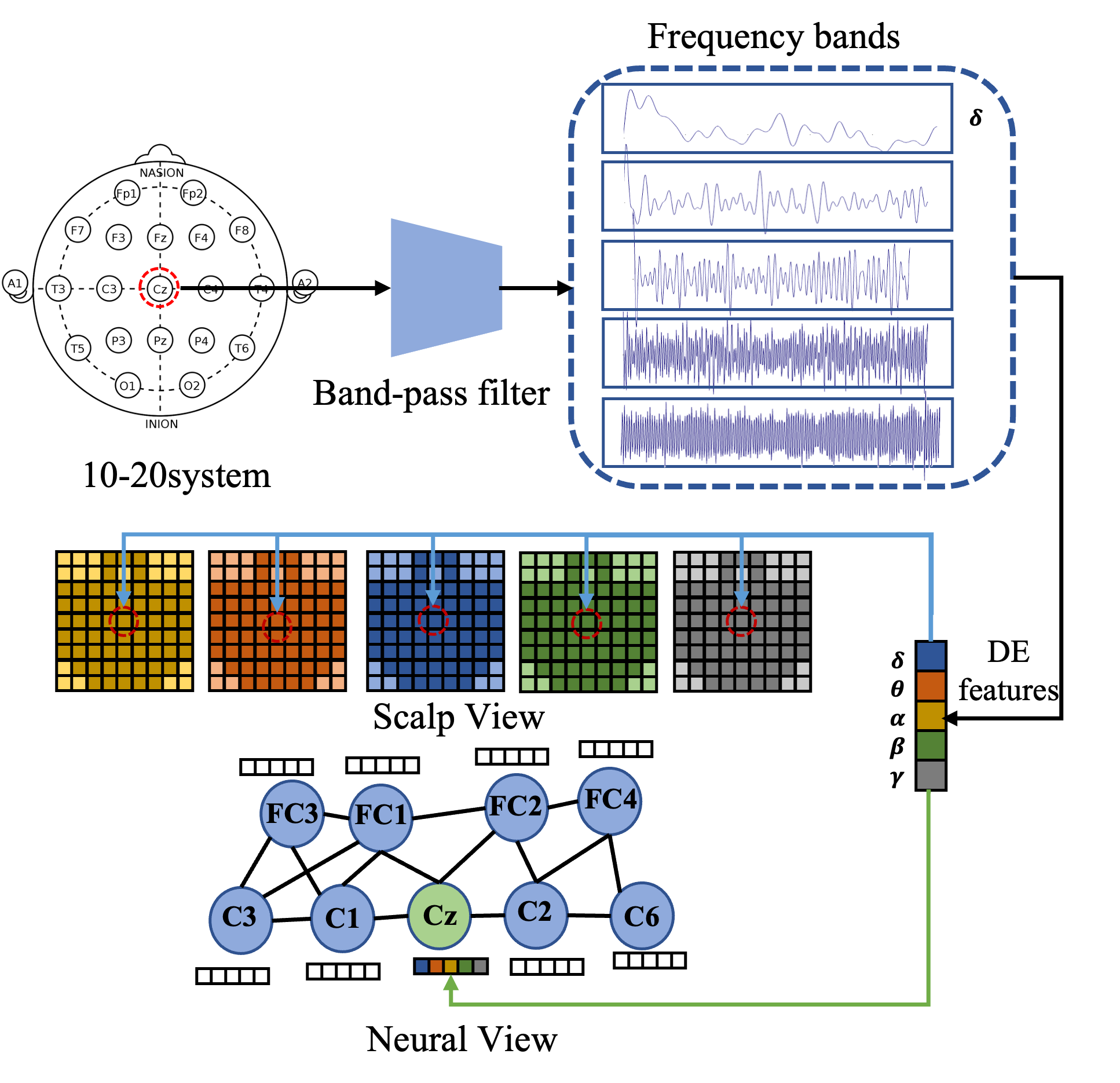}
  \caption{The construction of different views from EEG signals.}
  \label{fig4}
\end{figure}

\textbf{External scalp view} measures the spatial distribution of the voltage generated by the EEG signal on the scalp. The scalp view constructs the spatial matrix representing the spatial arrangement of the electrodes on the scalp according to the international standard 10-20 system. The spatial matrix is the designed two-dimensional matrix as the top-down projection of the scalp, where the rows and columns indicate the left-right positions and front-back positions of the electrodes on the scalp separately. The features of different electrodes are filled into the spatial matrix according to its spatial position, and zeros are used to fill the rest of the matrix, indicating no electrodes at the current position. Due to the electrode number in the general EEG-based brain tasks, the $9 \times 9$ spatial matrix can capture the spatial distribution of all electrodes. The preliminary features of EEG signals are transformed into the scalp view: the calculated differential entropy features of five frequency bands are filled into five different spatial matrices, which maps the  $M \in {\Bbb{R}}^{c \times 5}$ into the 3D scalp structure $SV \in {\Bbb{R}}^{ 9 \times 9 \times 5}$. Fig.3 shows the details of the scalp view construction. The generated EEG scalp view is the external spatial representation of neural activity in the brain, which describes the spatial correlations between electrodes.
 
\textbf{Internal neural topology view} measures the inner brain activity and the topological correlation between the neurons. The combination of different dipoles inside the brain generates the voltage measured by the electrodes placed on the scalp. Due to the theory that different pyramidal neurons (dipoles) in the brain are connected topologically, this paper adopts a graph structure to depict inner correlations between dipoles and constructs the neural topology view of EEG signals. Although the voltage collected by the electrode on the scalp is the superposition of multiple dipoles in different directions, the dipole perpendicular to the electrode's direction contributes the most to the voltage at this position. Based on the assumption that one dipole is responsible for the voltage on the specific position, the undirected graph $ \mathcal{G} = (\mathcal{V}, \mathcal{E})$ is constructed representing the internal neural topology view of the EEG signal. In the graph, each node $n \in \mathcal{V}$ represents a dipole corresponding to an electrode where the preliminary features of the electrode are formed as the node features. The $\mathcal{E}$ is the edge of the graph, where the connected edge between two nodes $(n_i,n_j) \in \mathcal{E}$. Two adjacent graph structures are used to simulate the topological neuron connections:  The nodes of adjacent spatial electrodes are connected with an edge, which assumes the localized and sparse connectivity of the neurons in the brain. 

\subsection{Learning of Two View's Representations}

As two different views are generated from the EEG signals, different base encoders are designed to learn neural representation from the external scalp and internal neural topology view separately. Despite many CNN-LSTM, CNN-Transformer structured methods have shown success in extracting EEG-related features, the complex spatial-temporal deep models are not suitable for learning representation from the generated two views, which may ignore the knowledge contained in different views and are useless in dealing with the simple structured data contains domain knowledge. In this part, simple structured networks are designed to extract effective representations from two views trained under knowledge-driven contrastive learning.

The shallow convolutional neural networks are designed to mine neural representations from external scalp view. The scalp view features $SV \in {\Bbb{R}}^{ 9 \times 9 \times 5}$ are similar to the three-channel image: the $9 \times 9 $ two-dimensional matrix can be analogized to the 2D images, while the number of frequency bands $5$ can be considered as the image channels (for example, RGB channels). Besides, the scalp view features exhibit a strong local spatial correlation similar to the image, which is well suited to capture local features through CNN. The shallow convolutional neural networks used designed in the base encoder are composed of multiple cascade CNN and pooling layer, where the CNN blocks with the kernel of $2 \times 2$ extract features from the local receptive field and pooling layer improve the translational invariance and robustness of the features. Due to the low spatial dimension of the scalp view features, the depth of conv-pooling is relatively shallow. The fully connected layers are placed at the end of the base encoder to map the features into the representation vector. Fig 3 shows the detailed structure of the base encoder designed for the scalp view of EEG signals.

The graph convolutional neural networks are designed as the base encoders for the inner neural topology view. The GCN fully utilizes the dipole topology structure to explore the inner correlation between the neurons, which captures the influence of neighboring and distant neurons on the current neuron until a state of equilibrium is reached. According to the undirected graph $\mathcal{G} $, the adjacency matrix $A \in {\Bbb{R}}^{c \times c}$ and the degree matrix $ D = \begin{matrix} \sum_{i=1}^c A_{ii} \end{matrix}$ are constructed representing the connectivity status of nodes. In the graph structure, the Laplacian operator replaces the spatial convolution kernel to conduct the spectral filters to extract the first-order neighbor features of nodes. The symmetric normalized Laplacian matrix are then calculated: $L^{sym}=\widetilde{D}^{-\frac{1}{2}} \widehat{A} \widetilde{D}^{-\frac{1}{2}}$, where $\widehat{A}= A+I $ is the normalized adjacency matrix that add the indentity matrix $I$, and $\widetilde{D} = \begin{matrix} \sum_{i=1}^c \widehat{A}_{ii} \end{matrix}$ is the normalized degree matrix. The spectral graph convolution is calculated as follows:

\begin{equation}
     g^{l+1}=\delta(\widetilde{D}^{-\frac{1}{2}} \widehat{A} \widetilde{D}^{-\frac{1}{2}} g^l \theta^l)
\end{equation}
where $g^{l}$ is the feature matrix at GCN layer $l$ that describes the representations of all the nodes in the graph (in the first layer is the input feature matrix), $\theta$ is the trainable variables and $\delta$ is the activation function. Multiple GCN layers are contained in the base encoder, with the fully connected layer mapping the features into the representation vectors. Fig. 3 shows the detailed structure of the base encoder designed for the inner neural topology view.

In summary, representations can be learned from two views through the designed base encoders separately. The representation generated from the scalp view is denoted as ${r_s}$ and the generated from the inner neural topology view is denoted as $r_t$, where both representations $r_t, r_s \in {\Bbb{R}}^{h}$.

\subsection{Hybrid Augmentation}
Different hybrid augmentation methods are used to generate augmented samples in different views. 
This paper uses four data augmentation methods to construct the positive and negative sample for the inner-view and cross-view contrastive learning: channel mask, spatial shuffle, frequency shuffle, and the hybrid method. The designed four augmentations are listed as follows: 

Mask augmentation method: The mask augmentation method is based on the idea of mask autoencoder \cite{r48}, which can reconstruct the detailed masking image information from the remaining image patches. The success of the masked autoencoder demonstrated that the part-masked samples still maintain the critical information and inner correlation. The mask augmentation method randomly masks different channels' features according to a mask rate $m_r$. Zeros are used to replace masked features. The mask-augmented samples contain correct spatial and topology structures but incomplete feature information.

Shuffle augmentation method: Inspired by the idea of the jigsaw pretext task \cite{r49} that has been widely used in computer vision to pre-train the model and extract effective image representations through restoring the shuffled image patches. In the shuffle augmentation method, two different shuffling strategies are designed: 1. The spatial shuffle strategy randomly disrupts the permutation of the electrodes. The augmented sample matrix $\widetilde{M} \in {\Bbb{R}}^{c \times 5}$ can be calculated as follows:
\begin{equation}
    \widetilde{M}=Rand(
    \begin{cases} 
    M_1=[{DE}_1,DE_2,...,DE_c]^T \\  
    M_2=[DE_1,DE_2,...,DE_{c-1}]^T \\
    ... \\
    M_{c!}=[DE_c,DE_{c-1},...,DE_1]^T
    \end{cases})_{c!}
\end{equation}
where $Rand()_{c!}$ represents the random choose function that chooses one augmented sample from the $c!$ electrodes permutations. 
2. The frequency shuffle strategy that randomly shuffles the features from different frequency bands. The calculated differential entropy features from five independent frequency bands are randomly shuffled within each electrode. The augmented samples $\widetilde{M}$ generated from frequency shuffle can be calculated as follows: 
\begin{equation}
    \widetilde{M}=
    \begin{bmatrix}
     Ch_1=Rand(f_1,f_2,...,f_5 )_{5!} \\
     Ch_2=Rand(f_1,f_2,...,f_5 )_{5!} \\
     ...\\\
     Ch_c=Rand(f_1,f_2,...,f_5)_{5!}
    \end{bmatrix} \in {\Bbb{R}}^{c \times 5}
\end{equation}
where the $Ch_i \in {\Bbb{R}}^{1 \times 5}$ represents the features of specific electrodes, and $f_1,f_2,...,f_5$ are the differential entropy calculated from five frequency bands. The shuffle augmented samples contain incorrect spatial and topology structures but complete feature information.

Besides, the hybrid augmentation method fuses both augmentation methods to generate samples that contain incorrect spatial and topology structure and incomplete feature information. In summary, three different augmentation methods formed the augmentation reservoir, and the $m$ augmentation samples are generated by the method randomly chosen from the reservoir. The generated augmentation samples $\widetilde{M_1},\widetilde{M_2},...,\widetilde{M_m}$ are transformed into two views and encoded into representations through different base encoders separately. In this way, the augmented representations in different from different views are mutually corresponding: In the scalp view, the representations of the augmented samples are denoted as $\{r_{sa}^1,r_{sa}^2,...,r_{sa}^m\} \in R_{sa}^*$, while in the neural view are denoted as $\{r_{ta}^1,r_{ta}^2,...,r_{ta}^m\} \in R_{ta}^*$. The next, combined contrastive methods are designed to self-supervised train the model by contrasting sample pairs inner and cross two views. 

\subsection{Inner View Contrastive Learning}
This paper conducts the inner view contrastive loss to help the encoders capture distinctive and invariant features. Traditional contrastive methods used in EEG signal \cite{r43} divided positive and negative pairs between different EEG time slices: the augmented samples from the same time slice were considered positive pairs, while the augmented samples from different time slices were considered negative pairs. However, EEG signals exhibit a strong temporal dependency, meaning the EEG time slices close in time contain highly similar abstract semantics. If samples from the close time slices are considered negative pairs and their information distance is pulled apart in the representation space, the invariant features contained in the samples would be ignored. The idea of the non-negative contrastive learning \cite{r50} (Barlow twins) is adopted in this paper to learn invariant features inner the view. Take the scalp view as an example, the generated representations of the augmented samples $\{r_{sa}^1,r_{sa}^2,...,r_{sa}^m\}$ are considered as positive sample pairs. Firstly, the cross-correlation matrix $C \in {\Bbb{R}}^{h \times h}$ are calculated between different representations along the batch dimension, which can be shown as follows:
\begin{equation}
    C_{i,j} \overset{\text{def}}{=}\sum_{p=1}^m {\sum_{q=p+1}^m {\frac{\begin{matrix}\sum_{b} {r_{sa,b}^p[i] r_{sa,b}^q[j]} \end{matrix}}{\sqrt{\sum_{b} {({r_{sa,b}^p[i]}^2)}} \sqrt{\sum_{b} {({r_{sa,b}^p[j]}^2)}}}
    }}
\end{equation}
where the $p$ and $q$ index the augmented representations in the scalp view, $b$ indexes the samples contained in the batch and the $i$ and $j$ index the dimensions of the representation generated by the base encoders. The calculated 
\begin{figure}[t]
  \centering
  \includegraphics[width=0.8\linewidth]{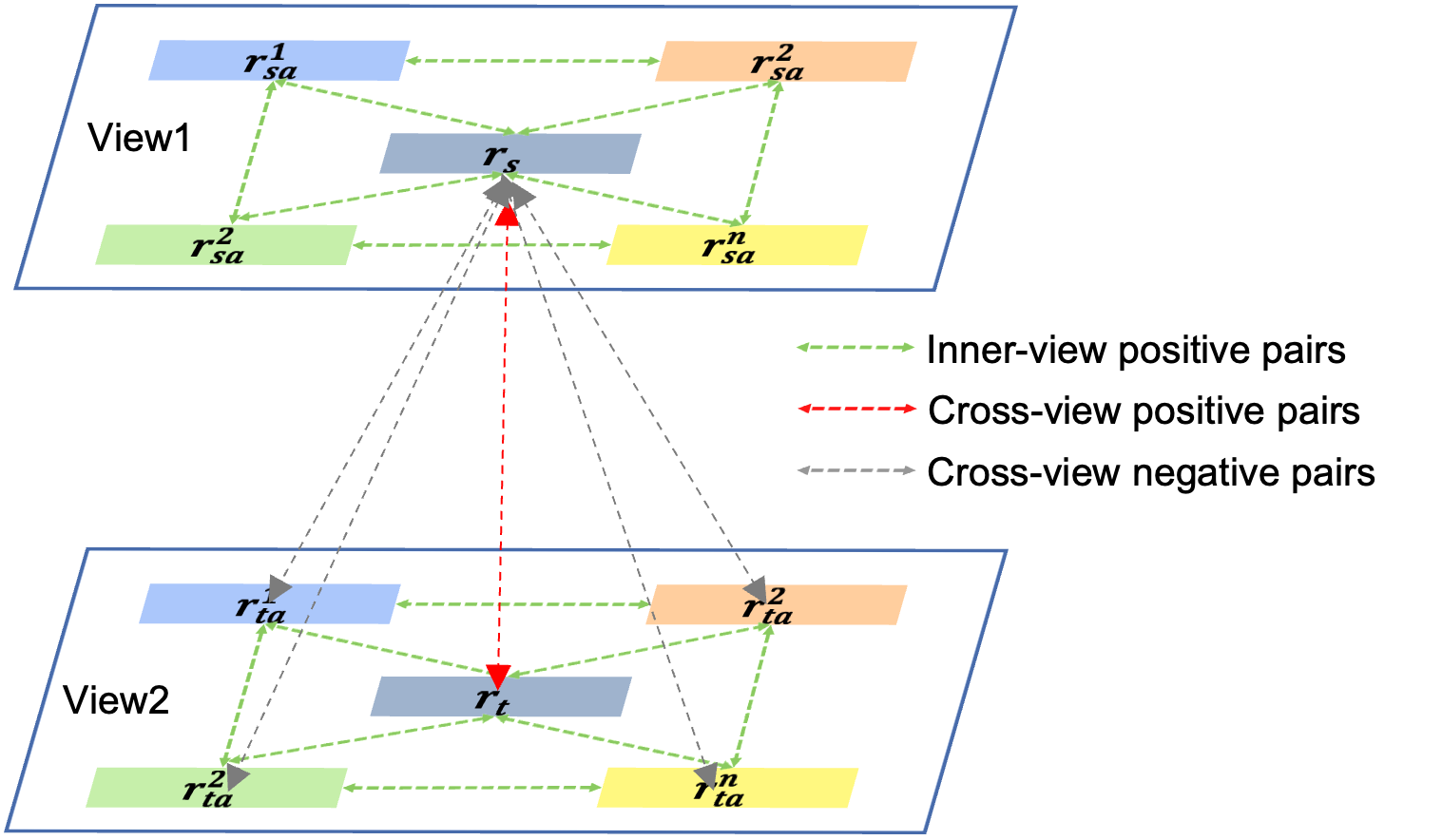}
  \caption{The construction of positive and negative pairs inner-view and cross-view.}
  \label{fig4}
\end{figure}
cross-correlation matrix measures the elements' correlation between the representations with the values in the matrix are limited to -1 (most anti-relevant) to 1 (most relevant). To extract the invariance features contained in multiple augmented samples, the Barlow twins loss \cite{r50} can be calculated:
\begin{equation}
    \mathcal{L}_{BT}^s=\sum_{i=1}^m {\frac{{(1-C_{i,i})}^2}{m}}+\sum_{i=1}^m{\sum_{j \ne i}^m {\frac{{C_{i,j}}^2}{m(m-1)}}} 
\end{equation}
where the $\mathcal{L}_{BT}^s$ is the Barlow twins (BT) loss calculated to train the encoders of scalp view. In the BT loss, the first term aims to set the diagonal values in the correlation matrix to 1. This term in the BT loss helps the model generate highly similar or relevant representations for different augmented samples, which can train the model to extract invariant features from discrepant information. The second term in the BT loss aims to set the other values in the correlation matrix to 0. This term in the loss function helps the model decouple the vector components of the representations to reduce the redundancy of representation components. Therefore, the Barlow twins loss can pre-train the model to preserve the invariant features between samples and ignore the information brought by the sample augmentation. Because the base encoders of two views are independent, the BT loss $\mathcal{L}_{BT}^s$ in the scalp view and $\mathcal{L}_{BT}^t$ in the neural view exist do not interfere
with each other and can be directly added to form the inner view contrastive loss, which is shown as follows:
\begin{equation}
    \mathcal{L}_{inner}=(\mathcal{L}_{BT}^s+\mathcal{L}_{BT}^t)/2
\end{equation}

\subsection{Cross View Contrastive learning}
Innovatively, the cross-view contrastive learning method is proposed in this paper, which incorporates the neural knowledge in the contrastive loss to extract complementary neural features between two views and generate the knowledge-driven neural representation. Firstly, the theory of the EEG generation are proposed as the prior neural knowledge \cite{r2}: the EEG 
\begin{figure}[t]
  \centering
  \includegraphics[width=0.6\linewidth]{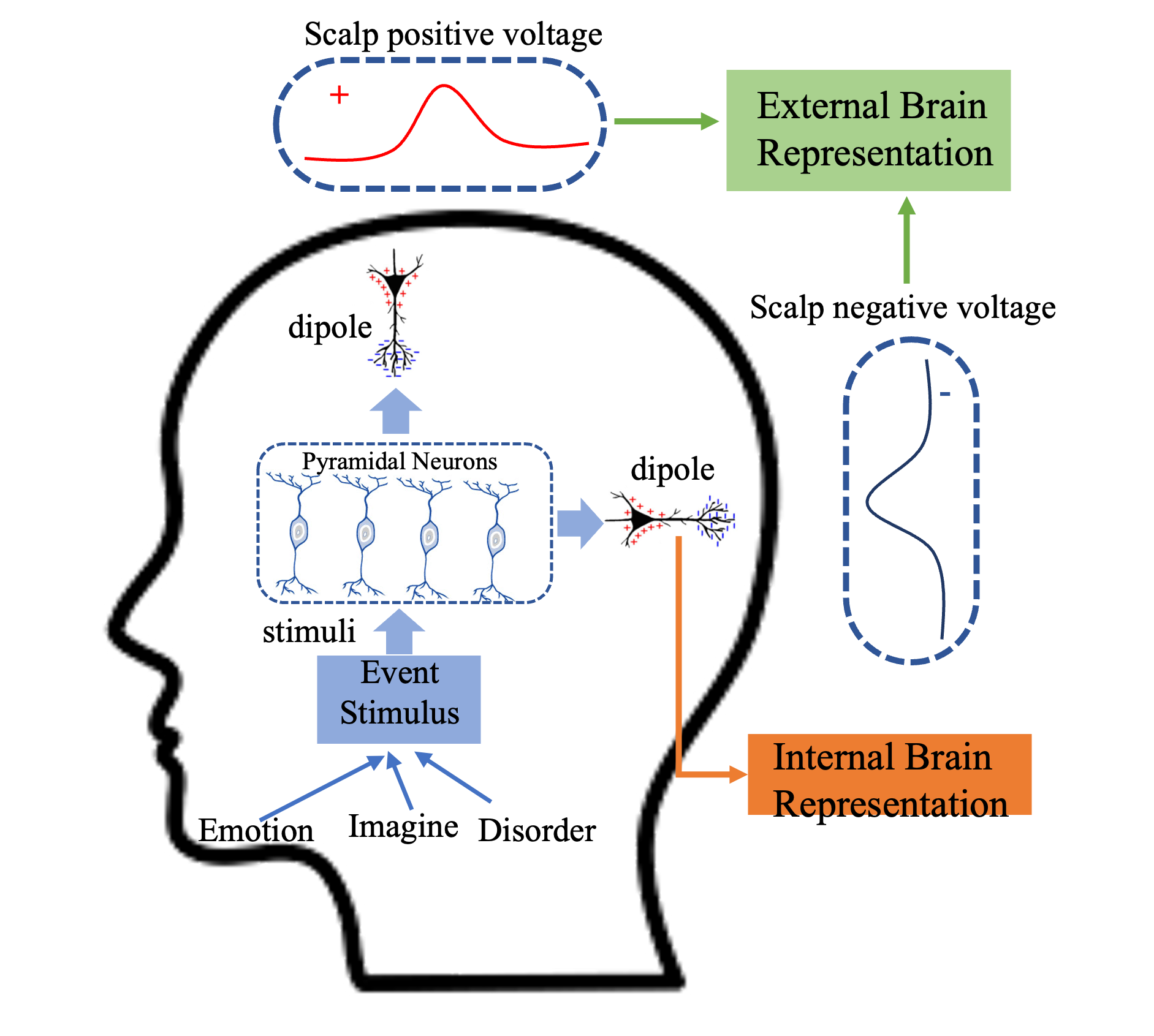}
  \caption{The schematic diagram of the neural source theory of EEG signals: generated from the inner dipoles.}
  \label{fig_new}
\end{figure}
signals are generated from the synchronize synaptic activity, which stimulates neuronal excitation and generates a negative extracellular voltage near the dendrites to changed the neurons into a dipole. The voltage generated by the dipole is transmitted to the scalp via capacity and volume conduction to be captured by electrodes as EEG signals, which are shown in Figure \ref{fig_new}. According to this prior knowledge, both scalp view and neural view are essentially neurological homologous and express similar information. Therefore, although different views describe the complementary internal and external brain activity separately, they express similar neural information in the abstract feature space representing the common knowledge of the neural activity. The knowledge-driven cross-view contrastive loss is designed to optimize the encoders to extract the hidden neural knowledge and generate the neural representations. The main idea of the knowledge-driven in this part is \textbf{"contrastive with same augmentations"}: Different augmented EEG samples are transformed into two views. Hence, the representations calculated in different views are one-to-one correspondence. The corresponding representation in different views constitutes the \textbf{positive pairs}: for example, the generated representations  $r_{sa}^1$ and $r_{ta}^1$, $r_{sa}^2$ and $r_{ta}^2$... are all considered as the positive sample pairs. By narrowing the distance between representations of the same augmented sample within different views in the information space, the model is pre-trained according to the neural information consistency theory to extract the neural knowledge expressed synergistically by two views. The \textbf{negative pairs} are made of the representations of different augmentation samples generated by different views: for example, the representation $r_{sa}^1 $ can form the negative pairs with the representations $\{r_{ta}^2,r_{ta}^3,..., r_{ta}^m\}$. By distancing the negative pair representations in the information space, the model learns complementary features that describe brain activity's internal and external manifestations from two views. Based on the InfoNCE loss \cite{r51}, the designed cross-view contrastive loss can be calculated as follows:

\begin{equation}
    \mathcal{L}_{cross}= -\frac{1}{\left\vert \mathcal{B} \right\vert}log(\frac{pair^+}{pair^++pair^-})
\end{equation}

\begin{equation}
    pair^+=\sum_{b \in \mathcal{B}} { \sum_{i=0}^m {exp(s(r_{sa,b}^i,r_{ta,b}^i)/ \tau )}}
\end{equation}

\begin{equation}
    pair^-=\sum_{b \in \mathcal{B}} { \sum_{i=0}^m { \sum_{j=i+1}^m{exp(s(r_{sa,b}^i,r_{ta,b}^j)/ \tau )}}}
\end{equation}
where $pair^+$ and $pair^-$ represents the cross-view positive and negative pairs, $\mathcal{B}$ is the sample batch and $\tau$ is the temperature parameter. The function $s()$ represents the cosine similarity to calculate the information distance between different representations. Through the cross-view contrastive loss, the encoders in two views are jointly trained to cooperate in extracting abstract neural knowledge and complementary features.

\subsection{Model Training}

With the inner view and cross view contrastive losses mentioned above, the model can be trained to extract effective neural representations for different EEG-based downstream brain tasks. For the input EEG signal, the generated representation $r_{sa}$ in the scalp view and the $r_{ta}$ in the neural view are concatenated into the neural representation $r_{f}= r_{sa} \oplus r_{ta}$, which can be further calculated by a task-specific decoder $\mathcal{H}(\cdot)$ (multiple linear layers with non-linear activation function) to accomplish specific classification tasks. Two different training modes optimize the model parameters based on the combined multiple losses.

\textbf{The pre-train and fine-tune mode} is the classic self-supervised in which the base model is first trained with no-label samples by the combined contrastive losses and fine-tuned tuned with limited labeled task samples by the classification loss. The pre-train and fine-tune mode can reduce the effect labeling issue in the EEG signal and generate more general representations. In the pre-train stage, the idea of uncertainty weighing \cite{r52} is combined to add some learnable noise parameters representing the uncertainty of different losses to fuse the inner and cross view contrastive losses, which can be calculated as follows: 
\begin{equation}
    \mathcal{L}_{pt}=\frac{1}{\sigma_s^2}\mathcal{L}_{inner}+\frac{1}{\sigma_t^2}\mathcal{L}_{cross}+log(\sigma_s)+log(\sigma_t)
\end{equation}
where the $\mathcal{L}_{pt}$ represents the combined losses, the $\sigma_s$ and $\sigma_t$ are the noise parameters trained in the loss function to control the relative weights of different contrastive losses. The combined loss pre-trains the model to extract general neural representations from non-label samples. In the fine-tuning stage, the parameters in the model are frozen, and the decoder $\mathcal{H}(\cdot)$ are fine-tuned by the task classification loss.  

\textbf{The joint-train mode} that combined the contrastive and the downstream task losses to comprehensively train the base encoder and decoders by the labeled task samples can reduce the overfitting and improve the task performance. The multi-task losses (contrastive task and downstream task) are combined to supervise train the model:
\begin{equation}
    \mathcal{L}_{jt}=\frac{1}{\sigma_{pt}^2} \mathcal{L}_{pt}+\frac{1}{\sigma_{ce}^2} \mathcal{L}_{ce}(y,\mathcal{H}(re))+log(\sigma_{pt}\sigma_{ce})
\end{equation}
where the $\mathcal{L}_{ce}(\cdot,\cdot)$ is the cross-entropy measures the loss between predicted results $\mathcal{L}(re)$ and the label $y$ on downstream tasks.

\section{Experiments}
To evaluate the performance of the proposed method, experiments were conducted on several benchmark datasets for different EEG-based downstream tasks. The experiments aim to answer the following research questions:

\textbf{RQ1:} Compared with the existing SOTA EEG representation method, can the proposed method perform better on different downstream tasks?

\textbf{RQ2:} Can the proposed representation method perform well in the few-label scenarios? 

\textbf{RQ3:} Are the main components (the inner view contrastive and cross view contrastive) all performing well in the whole model?

\textbf{RQ4:} Is the proposed model sensitive to hyper-parameters and the backbone structures?

\subsection{Benchmark Dataset}
To verify the proposed method, three different EEG-based benchmark datasets are used to evaluate the performance of the proposed method: the SEED dataset, the MMI dataset, and the CHB-MIT dataset. Different datasets contain different EEG-based downstream tasks, including emotion generation, motor imagery, and seizure analysis. The details of the benchmark dataset are listed as follows:

\textbf{SEED}\cite{r53}: The SEED dataset is the emotional EEG dataset that elicits three kinds of emotions (positive, neutral, negative) by video stimulus. This dataset contains 15 subjects, and the signals are collected by 62 electrodes with sampling rates of 200Hz.

\textbf{MMI}\cite{r54}: The MMI is a motor imagery dataset that collected EEG data from the subjects completing three motor imagery tasks: resting baseline, imagining opening and closing the left fist, and imagining opening and closing the right fist). This dataset contains 109 subjects and the EEG signals are collected by 64 electrodes with the sampling rates of 160Hz.

\textbf{CHB-MIT}\cite{r55}: The CHB-MIT EEG-based disorder dataset collected the EEG signal of epilepsy patients in the onset and normal states. This dataset contains 22 pediatric subjects with intractable seizures, and EEG signals are collected by 23 electrodes with sampling rates of 256Hz.

\subsection{Experiment Settings}
\subsubsection{Baseline methods}
Different SOTA algorithms are compared in the experiments to evaluate the proposed method's performance: classical self-supervised learning frameworks used in CV and NLP, EEG-based self-supervised methods, and some traditional supervised models as the supervised baseline. The classical self-supervised learning frameworks are listed as follows:

(1) \textit{Barlow twins}\cite{r50}: Barlow twins is the typical non-negative contrastive framework that learns invariant features from the augmented samples. The well-designed Barlow twins loss is used to optimize the model.

(2) \textit{BYOL}\cite{r56}: BYOL is a non-negative contrastive structure, which conducted the target and online networks to interact with each other and update alternately. The BYOL learns to extract representations from positive pairs directly.

(3) \textit{MOCO}\cite{r57}: MOCO is a general contrastive structure that contains positive pairs and negative pairs. The MOCO adopts a momentum encoder to introduce more negative samples into contrastive learning.

(4) \textit{SimCLR}\cite{r51}: SimCLR is the traditional contrastive learning framework. The augmented samples are divided into positive and negative pairs with the infoNCE loss to optimize the model and generate representations from augmented samples.

(5) \textit{SimSiam}\cite{r58}: SimSiam is the non-negative contrastive learning framework that adopts the siamese network structure. The SimSiam method uses a symmetrized loss to learn invariant features between two augmented samples.

(6) \textit{Jigsaw}\cite{r49}: The Jigsaw is the common self-supervised pretext task in the visual field. The task is to recover the randomly shuffled image patches. In the EEG field, the task is translated into recovering the shuffled electrodes.

The EEG-based SOTA self-supervised methods are listed:

(1) \textit{GMSS}\cite{r19}: GMSS is the graph-based hybrid self-supervised learning method. The GMSS combines the spatial shuffle, frequency shuffle and contrastive tasks to form the multi-task SSL loss and jointly optimize the model.

(2) \textit{Trans-SSL}\cite{r39}: The trans-SSL is the CNN-based EEG representation method, which designs the pretext task to recognize the transformation added to the raw EEG signal. 

(3) \textit{TS-SSL}\cite{r59}: The TS-SSL method is the CNN-based EEG representation method, which conducts the pretext task to accurately predict the order of shuffled temporal EEG signals to help the model learn effective representations from unlabeled samples.

(4) \textit{Sim-EEG}\cite{r43}: The Sim-EEG is the extension of the Simclr in the EEG field. The data augmentation method in the Sim-EEG is replaced by the signal processing method, like the DC shift, the band-stop filter, and the additive noise.

Besides, the traditional supervised methods are listed:

(1) \textit{4DCRNN}\cite{r60}: The 4DCRNN is the supervised method that constructs the DE-based spatial matrix from five frequency bands forming the 4D structure, with the CNN-LSTM structure to extract the task-related features.

(2) \textit{ACRNN}\cite{r27}: The ACRNN is an end-to-end EEG analysis framework that combines CNN, LSTM, and channel-wise attention to fuse the spatial-temporal features in the EEG signal for specific EEG-based tasks.

\subsubsection{Model Implementations}

Different settings were applied to the proposed method in the experiment according to the data format. The batch size was 256 and the learning rate was set to 0.01 with Adam optimizer to optimize the model. For the model structure, the number of the augmented samples generated in each view was set to 3. The augmented number means that for any EEG sample, the number of positive pairs in the inner-view contrastive learning was 3. while the number of positive and negative pairs in the cross-view contrastive learning was 3. The dimension of the generated representation was set to 128. For the SEED dataset, the recorded signals were divided into 3s samples with overlapping, so we can get a total number of 2700 samples for each subject of three emotions; For the MMI dataset, each motor imagery task lasted for 4s, so it can straightforward to divide the trail into 4s samples; For the CHB-MIT dataset, the signals were divided into the 1-second samples for detecting short-term seizure. 54,000 samples can be divided for all subjects with the seizures extracted and mixed with non-seizure samples. 

Two different training modes are used to evaluate the performance of the models: 1. the pre-train mode, the model is self-supervised trained on the non-label training samples and then fine-tuned with labeled samples; 2. the joint-train mode, the model combines contrastive losses and the task loss are fused as multi-task learning to joint supervised train the model on the labeled training samples. The compared baseline methods are adjusted for the experiments: for the classical self-supervised frameworks, the data augmentation methods for those models were set to the mask and shuffle consistent with the proposed methods, and the jigsaw problem changed from image patch jigsaw to electrodes jigsaw, the encoders were set to the cascade shallow CNN and the projection head were set to the multi-layer fully connected layers. For the supervised methods, the time slice length is set to 0.2s to extract short-term temporal dependencies through the RNN and LSTM. The rest of the settings followed the original settings of the baseline algorithms.

\subsection{Performance Comparison (RQ1)}
The performance of the proposed method was first verified through comparison experiments. This paper compared different methods on three downstream tasks: 
\begin{table*}[htbp]
\small
\centering
\caption{The comparison experimental results on three widely used downstream task benchmark datasets. }
\begin{tabular}{c|cc|cc|cc} \hline 
\multirow{2}{*}{\textbf{Model}} & \multicolumn{2}{c|}{\textbf{SEED Dataset}} & \multicolumn{2}{c|}{\textbf{MMI dataset}} &  \multicolumn{2}{c}{\textbf{CHB-MIT}} \\ 

 & \textbf{Pre-training} & \textbf{Joint-training} & \textbf{Pre-training} & \textbf{Joint-training}  & \textbf{Pre-training} & \textbf{Joint-training}\\ \hline \centering
Barlow twins(2021) & 80.17$\%$ & 84.11$\%$ & 91.84$\%$ & 92.18$\%$ & 96.71$\%$ & 98.19$\%$ \\ 
SimSiam(2021)       & 79.29$\%$ & 82.17$\%$ & 90.06$\%$ & 91.07$\%$ & 96.89$\%$ & 98.19$\%$ \\ 
BYOL(2020)          & \underline{81.34$\%$} & \underline{86.75$\%$} & 91.78$\%$ & 92.85$\%$ & \underline{97.24$\%$} & 97.96$\%$ \\ 
SimClr(2020)        & 81.07$\%$ & 82.93$\%$ & \underline{93.67$\%$} & \underline{94.54$\%$} & 97.16$\%$ & \underline{97.99$\%$} \\ 
MOCO(2020)          & 79.16$\%$ & 80.21$\%$ & 92.15$\%$ & 92.40$\%$ & 96.53$\%$ & 98.24$\%$ \\ 
Jigsaw(2016)        & 74.65$\%$ & 78.16$\%$ & 85.74$\%$ & 90.07$\%$ & 97.12$\%$ & 97.41$\%$ \\ 
\hline
GMSS(2022) & \underline{89.17$\%$} & \underline{91.78$\%$} & \underline{93.27$\%$} & \underline{94.80$\%$} & \underline{98.19$\%$} & \underline{98.76$\%$} \\ 
Trans-SSL(2021) & 80.93$\%$ & 86.71$\%$ & 90.21$\%$ & 91.32$\%$ & 96.57$\%$ & 97.97$\%$ \\ 
TS-SSL(2021) & 76.01$\%$ & 83.59$\%$ & 91.12$\%$ & 92.07$\%$ & 96.82$\%$ & 98.59$\%$ \\ 
Sim-EEG(2021) & 81.45$\%$ & 82.17$\%$ & 91.25$\%$ & 92.76$\%$ & 95.74$\%$ & 97.83$\%$ \\  
\hline
4DCRNN(2022) & - & 80.76$\%$ & - & 91.26$\%$ & - & 98.66$\%$ \\ 
ACRNN(2020)  & - & \underline{83.36$\%$} & - & \underline{93.18$\%$} & - & \underline{98.83$\%$} \\  
CNN & - & 73.45$\%$ & - & 89.75$\%$ & - & 91.07$\%$ \\
LSTM & - & 76.19$\%$& - & 85.32$\%$ & - & 90.35$\%$ \\
MLP & - & 71.73$\%$ & - & 82.79$\%$ & - & 90.87$\%$ \\
SVM & - & 73.85$\%$ & - & 80.52$\%$ & - & 92.18$\%$\\
\hline
Our method & \textbf{93.58$\%$} & \textbf{97.26$\%$}  & \textbf{94.30$\%$} & \textbf{95.33$\%$}  & \textbf{98.94$\%$}  & \textbf{99.47$\%$}  \\
Improved & 4.41$\%$ & 5.48$\%$ & 0.63$\%$ & 0.53$\%$& 0.80$\%$ &  0.64$\%$ \\
\hline
\end{tabular}
\label{table_1}
\end{table*}
emotion recognition task, motor imagery task, and epilepsy recognition task. In this experiments, the downstream task classification accuracy was used to evaluate the performance of different methods. Table  \ref{table_1} demonstrates the result of the comparison experiments. In Table \ref{table_1}, the underlined value indicates the optimal result compared with the same type of algorithms in the same experiment, the bolded value represents the global optimal result. 

From the results in Table \ref{table_1}, the proposed method performed best on three EEG-based downstream tasks. For the triple classification emotion recognition task, the proposed method improved the classification accuracy by 4.41$\%$ and 5.48$\%$ on pre-training and joint-training modes separately compared with the SOTA method. The result indicates that the KDC2 method can significantly enhance the ability to extract emotion-related brain neural features. Besides, the proposed KDC2 method also outperforms the SOTA method on other tasks: the classification accuracy improved by about 0.53$\%$ and 0.64$\%$ through pre-training modes on MMI and CHB-MIT separately, while the accuracy improved by about 0.63$\%$ and 0.80$\%$ through joint-training. The improvements in the motor imagery and the epilepsy recognition tasks are not as significant as the emotion task, possibly because the emotion recognition task is more complex than the other task. Unlike other tasks with objective descriptions (such as the imagination process of motor imagery and the epileptic seizure state), subjects have different subjective standards for emotions, and there are significant differences in the collected emotional EEG signals elicited by emotional videos. Traditional methods have difficulties extracting emotion-related features, but the proposed KDC2 method can effectively improve the classification accuracy, demonstrating the superior performance of the proposed cross-view contrastive learning framework to learn emotion-related neural representations. On the joint-training modes, by combining the task loss with the contrastive loss proposed in the KDC2 method, the classification accuracy improved, indicated that the joint-training modes can
\begin{figure}[t]
\centering
\subfigure[Pre-train on SEED]{
\begin{minipage}[t]{0.49\linewidth}
\centering
\includegraphics[width=1\linewidth]{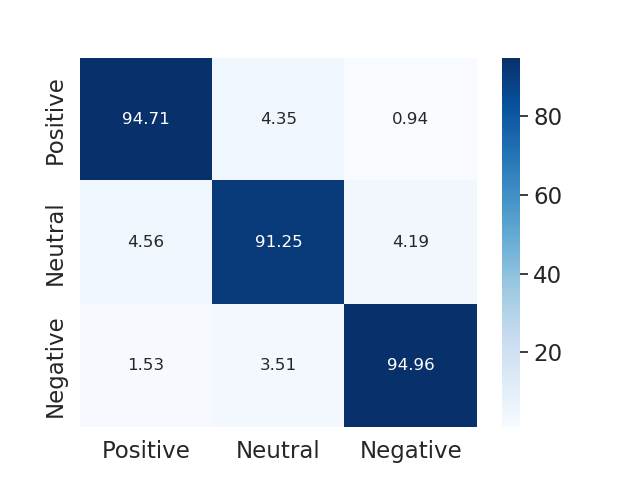}
\end{minipage}%
}%
\subfigure[Joint-train on SEED]{
\begin{minipage}[t]{0.49\linewidth}
\centering
\includegraphics[width=1\linewidth]{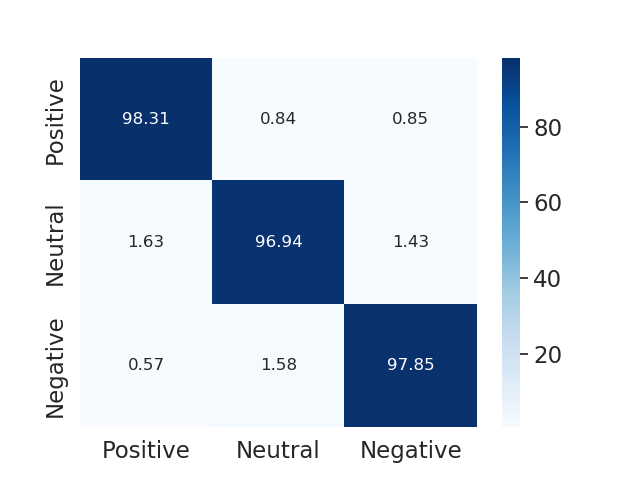}
%\caption{fig2}
\end{minipage}%
}%
\\
\subfigure[Pre-train on MMI]{
\begin{minipage}[t]{0.49\linewidth}
\centering
\includegraphics[width=1\linewidth]{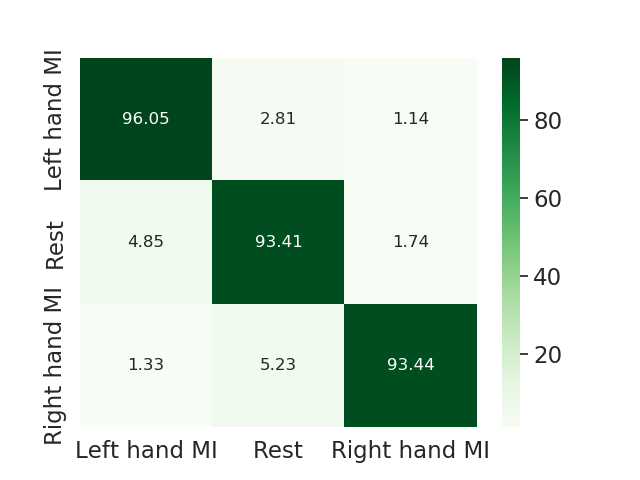}
\end{minipage}%
}%
\subfigure[Joint-train on MMI]{
\begin{minipage}[t]{0.49\linewidth}
\centering
\includegraphics[width=1\linewidth]{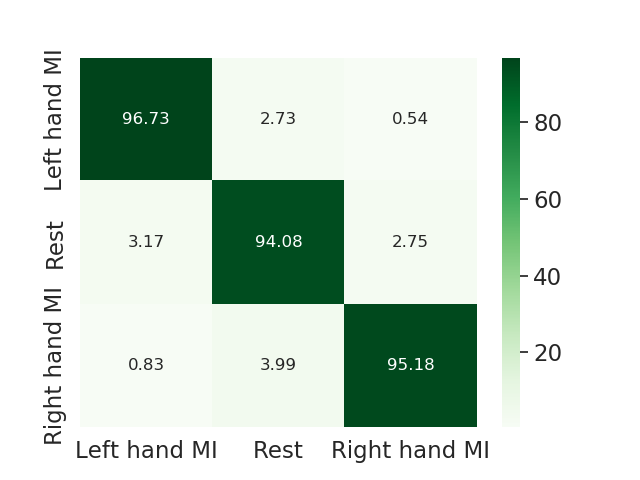}
%\caption{fig2}
\end{minipage}%
}%
\centering
\caption{The confusion matrix of the proposed method on different datasets through pre-train and joint-train. }
\label{confusion matrix}
\end{figure}
do mitigate the overfitting problem that widely occurs in single-task learning. On the pre-training modes, the results indicated that the neural representation generated by the KDC through self-supervised training is highly effective for various downstream tasks. So the RQ1 can be answered: The proposed knowledge-driven contrastive method can extract more general neural features that outperform the SOTA on both downstream tasks.

\begin{table*}[t]
\small
\centering
\caption{The few label experiments results on three downstream tasks.}
\resizebox{\linewidth}{!}{
\begin{tabular}{c|cccc|cccc|cccc} \hline 
\multirow{2}{*}{\textbf{Model}} & \multicolumn{4}{c|}{\textbf{SEED Dataset (percentage of labels)}} & \multicolumn{4}{c|}{\textbf{MMI dataset (percentage of labels)}} &  \multicolumn{4}{c}{\textbf{CHB-MIT (percentage of labels)}} \\ 

 & \textbf{75$\%$} & \textbf{50$\%$} &\textbf{25$\%$} &\textbf{1$\%$}  & \textbf{75$\%$} & \textbf{50$\%$} &\textbf{25$\%$}  &\textbf{1$\%$}  & \textbf{75$\%$} & \textbf{50$\%$} &\textbf{25$\%$}  &\textbf{1$\%$}\\ \hline \centering
Barlow twins(2021) & 78.31$\%$ & 75.29$\%$ & 72.75$\%$ & 65.41$\%$ & 90.16$\%$ & 88.30$\%$ & 84.15$\%$ & 81.03$\%$ &  96.04$\%$ & 95.74$\%$ & 93.71$\%$ & 91.06$\%$ \\ 
SimSiam(2021)    & 77.34$\%$ & 75.10$\%$ & 71.07$\%$  & 64.13$\%$ & 88.71$\%$  &86.35$\%$ & 82.97$\%$ & 80.62$\%$& 96.03$\%$ & 95.42$\%$ & 93.17$\%$ & 89.73$\%$\\ 
BYOL(2020)  &  80.53$\%$ &  76.34$\%$ & 72.19$\%$  & 67.14$\%$ & 89.17$\%$ & 86.35$\%$ & 83.77$\%$ & 82.05$\%$ & 95.89$\%$ & 95.07$\%$ & 93.12$\%$ & 91.86$\%$  \\ 
SimClr(2020)  &  80.32$\%$ & 75.79$\%$ & 72.05$\%$ & 66.98$\%$  & 91.29$\%$ & 89.07$\%$ & 85.36$\%$ & 83.70$\%$ & 96.83$\%$ & 94.51$\%$ & 92.18$\%$ & 91.79$\%$   \\ 
MOCO(2020)     & 76.51$\%$ & 73.04$\%$  & 71.62$\%$  & 64.03$\%$ & 90.82$\%$ & 88.63$\%$ & 84.01$\%$ & 81.85$\%$ & 95.32$\%$ & 93.81 $\%$ & 91.72$\%$ & 89.90$\%$\\ 
Jigsaw(2016)    & 71.05$\%$ & 67.53$\%$ & 63.20$\%$ & 59.12$\%$  & 83.15$\%$ &80.92$\%$ & 77.16$\%$ & 74.03$\%$ & 96.41$\%$ & 93.98$\%$ & 91.05$\%$ & 88.95$\%$  \\ 
GMSS(2022) &  84.32$\%$ & 81.05$\%$ & 76.35$\%$ & 74.91$\%$ & 91.08$\%$ & 88.65$\%$ & 86.13$\%$ & 84.37$\%$ & 97.03$\%$ & \textbf{96.47$\%$} & \textbf{94.71$\%$} & \textbf{93.02$\%$} \\ 
Trans-SSL(2021) & 77.85$\%$ & 74.01$\%$ & 71.53$\%$ & 67.30$\%$ &88.64$\%$ & 85.48$\%$ &82.95$\%$ & 80.11$\%$ & 95.31$\%$ & 93.71$\%$ & 92.26$\%$ & 90.04$\%$  \\ 
TS-SSL(2021) & 72.85$\%$ & 68.13$\%$ & 64.75$\%$ & 62.84$\%$ & 89.53$\%$ & 85.30$\%$ & 82.73$\%$ & 79.67$\%$ & 95.43$\%$ & 93.77$\%$ & 91.42$\%$ & 89.62$\%$ \\
Sim-EEG(2021) & 79.17$\%$ & 76.09$\%$ & 72.83$\%$ & 68.11$\%$ & 90.06$\%$ & 87.87$\%$ & 84.72$\%$ & 82.06$\%$  & 94.13$\%$  & 92.04$\%$  &90.57$\%$ & 88.65$\%$ \\  
\hline
Our method & \textbf{92.07$\%$} & \textbf{89.15$\%$}& \textbf{86.53$\%$} & \textbf{84.75$\%$} &\textbf{92.31$\%$} & \textbf{90.77$\%$}  & \textbf{89.13$\%$} & \textbf{86.25$\%$}  & \textbf{97.65$\%$}  & 95.33$\%$ &92.17$\%$ & 91.02$\%$  \\
\hline
\end{tabular}
}
\label{table_2}
\end{table*}
We also analyzed the experimental results for the SOTA comparison methods. The first 6 lines in Table \ref{table_1} are the classical self-supervised frameworks widely used in computer vision. The BYOL and the SimCLR method achieve the best results on different tasks, the mask and shuffle methods widely used in computer vision are chosen as the 
augmentation methods in the self-supervised methods. The next 4 lines in Table \ref{table_1} are the SOTA SSL framework designed for EEG. The comprehensive methods GMSS achieved best on both tasks, this method designed different task heads to solve the frequency jigsaw, spatial jigsaw, and contrastive task, which indicates that the combined multiple related pretext tasks can extract more efficient and general features than the single task SSL framework in different EEG tasks. Moreover, the traditional supervised methods were also compared in the experiments, the self-attention combined CNN-LSTM framework performed well with sufficient labels, and the supervised spatial-temporal structure can approximate complex self-supervised frameworks on certain simple tasks (e.g., binary epilepsy classification).

Besides, Figure \ref{confusion matrix} demonstrated the confusion matrix of the proposed method. Because epilepsy recognition is a binary classification task, the confusion matrix of this task was not given in the experiments. For the emotion recognition task, the confusion matrix demonstrated that the positive emotions are easiest to classify by the proposed method, while the neutral emotions are easily misclassified. For the motor imagery task, the rest state is most likely confused with other imagery states(left hand, right hand). It can be concluded that the fuzzy intermediate states of different tasks are easily confused with other states, influencing task performance and recognition accuracy.

\subsection{Few-label experiment(RQ2)}

For research question 2, the experiments were designed to show the performance of the proposed method in the few label scenarios. This paper conducted the few label scenarios for all three EEG-based downstream tasks: We randomly reduced the labeled samples for fine-tuning to 75$\%$, 50$\%$, 25$\%$, 1$\%$ of the original samples, which simulates the scarcity of labeled samples in downstream task fine-tuning. In different EEG dataset, the pre-training stage are not influenced by the number of labels (in the pre-training stage, the models are trained self-supervised to learn the neural representation), so the few-label experiments only verify the fine-tuning adaptation performance of the generated neural representation in the label-less downstream task to show whether the proposed representation method can alleviate or solve, the labeling issued occurred in the EEG field. Table 2 shows the result of the few-label experiments.

In Table \ref{table_2}, all the listed self-supervised methods were pre-trained on the non-label task samples and then fine-tuned through the labeled task samples. The fine-tuned labeled samples are randomly reduced to 75$\%$, 50$\%$, 25$\%$, and 1$\%$ of the original samples to simulate the few-label downstream task scenarios separately. For the emotion recognition and the motor imagery task, the models' performance on downstream tasks decreased significantly with the number of labeled task samples continuing to decrease. For example, if the percentage of labeled samples for task fine-tuning decreases from 100$\%$ to 1$\%$, the existing representation methods witnessed a decline in fine-tuned accuracy of the emotion recognition task, which reduced from more than 80$\%$ to 65$\%$. Besides, the recognition accuracy in the motor imagery task reduced from more than 91$\%$ to about 80$\%$. However, the proposed cross-view contrastive method performed better on the few-label scenarios than the SOTA representation methods. When the percentage of labeled samples reduce to 1$\%$, the proposed methods can still maintain the accuracy of 84.75$\%$ on the three-class emotion recognition task and 86.25$\%$ on the three-class motor imagery task. Compared to the result of 75$\%$, the accuracy decreased by less than 8$\%$. The results indicate that the generated neural representation can adopt a few label EEG-based tasks, which validates the effectiveness of the proposed method in mining the general EEG neural features. The RQ2 can be answered by the result of few-label experiments, the proposed method performed well in the few-label scenarios, which can further help to mitigate labeling issues in the EEG field: to pre-train the model without labels and only fine-tuned with few labeled task samples to achieve efficient task model but only require limited labels. Besides, reducing the number of labeled samples for fine-tuning has no significant impact on the epilepsy detection task in the CHB-MIT dataset, which may be because the EEG signals of epileptic and non-epileptic states are distinctly different and highly differentiable.

\subsection{Components Ablation Experiments(RQ3)}

Besides, the components ablation experiments were designed to verify the rationality of the model structure. The effect of different architectures of the proposed method was tested in the experiments: 1. Test the effect of the whole self-supervised learning framework proposed in this paper, the experiment supervised training the model through the labeled samples without any self-supervised tasks, which were named as the \textbf{Model-NP}, 
\begin{figure*}[t]
\centering
\subfigure[Dimension experiments]{
\begin{minipage}[t]{0.25\linewidth}
\centering
\includegraphics[width=1\linewidth]{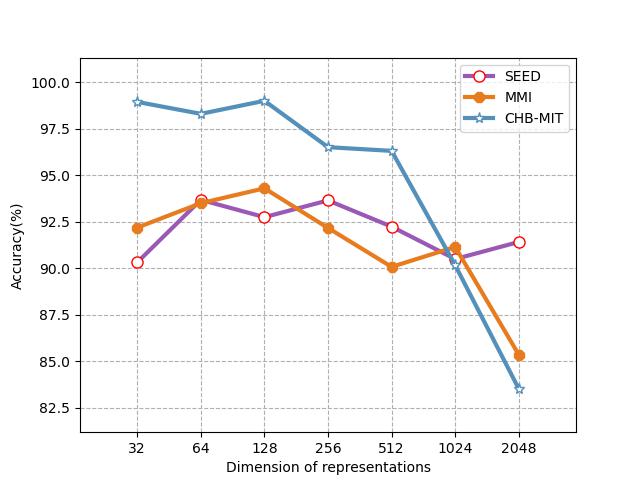}
\end{minipage}%
}%
\subfigure[Batch-size experiments]{
\begin{minipage}[t]{0.25\linewidth}
\centering
\includegraphics[width=1\linewidth]{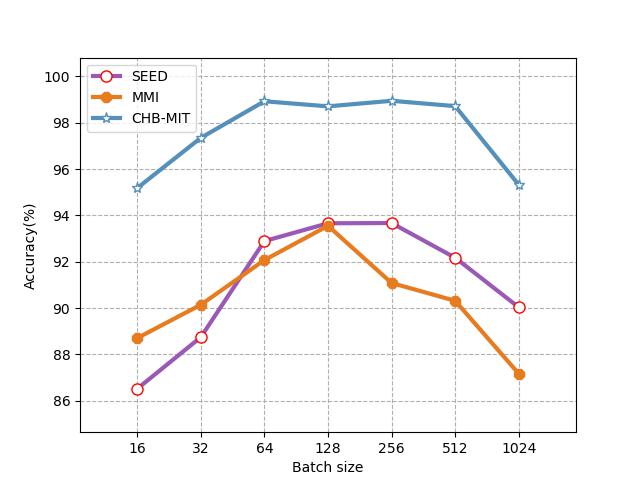}
%\caption{fig2}
\end{minipage}%
}%
\subfigure[Number of augmentations]{
\begin{minipage}[t]{0.25\linewidth}
\centering
\includegraphics[width=\linewidth]{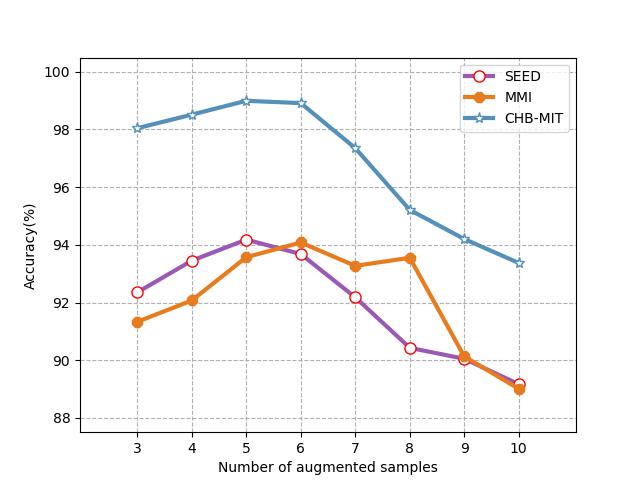}
%\caption{fig2}
\end{minipage}%
}%
\subfigure[Augmentation methods]{
\begin{minipage}[t]{0.25\linewidth}
\centering
\includegraphics[width=1\linewidth]{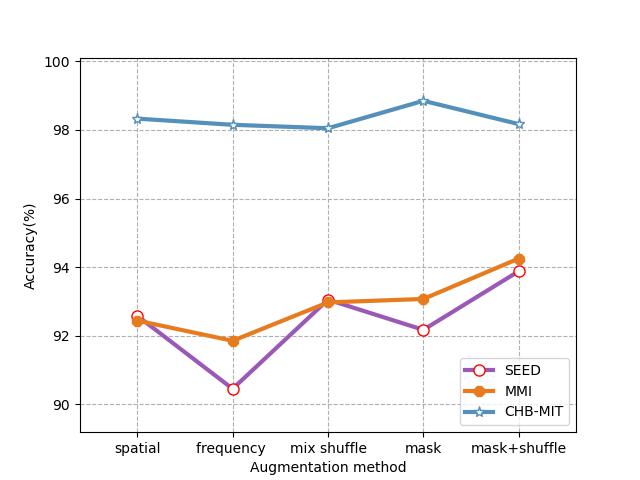}
%\caption{fig2}
\end{minipage}%
}%
\centering
\caption{The results of the parameter experiments (RQ4).The experimental results are displayed in a line graph: (a) demonstrates the dimension ablation experiments; (b) demonstrates the batch-size ablation experiments; (c) demonstrates the influence of the number of augmentation samples in the contrastive framework; (d) demonstrates the influence of the different augmentation methods used in the framework.}
\label{RQ4}
\end{figure*}which is consistent with the theory of brain region asymmetry \cite{r19}. 2. Test the effect of the inner-view and cross-view contrastive learning framework. In the experiments, the model contains only inner-view or cross-view contrastive tasks were tested, where the \textbf{Model-IN(T)}, \textbf{Model-IN(S)}, and \textbf{Model-IN(TS)} represents the model contained only the inner-view contrastive loss in the topology view, scalp view and both views respectively, and \textbf{Model-Co} is the model contained only the cross-view contrastive loss. 3. Test the effect of the features extracted from different views. In the experiment, the self-supervised losses composed of combined contrastive losses were not changed, but only the features from one view were used to generate representations. The \textbf{Model-P} and \textbf{Model-S} represents the models for generating neural representations using only spatial and topological perspectives, respectively. 

Table \ref{table 3} can answer the RQ3 that all the main components worked well in the model. By applying the combined contrastive losses to pre-train the model, the performance on different EEG-based downstream tasks improved by 9$\%$, 5$\%$, and 3$\%$, respectively. Both inner view and cross view contrastive loss functions are helpful to extract general features and models performed poorly when part of the contrastive loss is missing. Besides, the results indicate that the features mined by cross-view contrastive learning are more effective than inner-view contrastive learning, which can achieve better recognition accuracy. The experiments also proved that the features from different views play important roles in the model, when neural representations no longer incorpoßrate the features from the scalp view or topology view, the task performance drops significantly. All components (structures and contrastive losses) worked well in different tasks.

\subsection{Parameter Ablation Experiments(RQ4)}

The final experiments are the parameter ablation experiments to test the parameter sensitivity of the proposed method. In the experiment, several critical parameters are the focus of this paper: 
\begin{table}[t]
\caption{The ablation experiments of the backbone network (sensitivity to backbone networks) on different datasets.}
\begin{center}
\label{table 3}
\begin{tabular}{ c c c c }
\hline
\textbf{Model} & \textbf{SEED} & \textbf{MMI} &\textbf{CHB-MIT} \\
\hline
CNN+Cheb& 93.08$\%$ & 93.86$\%$ & 98.35$\%$ \\ 
CNN+GAT & 92.78$\%$ & 94.25$\%$ & 98.41$\%$ \\
FC+GCN & 90.57$\%$ & 93.17$\%$ &93.12$\%$ \\
FC+Cheb & 89.64$\%$ & 92.06$\%$ & 92.05$\%$ \\
FC+GAT  & 88.04$\%$ & 93.20$\%$ & 93.54$\%$ \\
\hline
\textbf{CNN+GCN} & \textbf{93.58$\%$} & \textbf{94.30$\%$} &\textbf{98.94$\%$}  \\
\hline
\end{tabular}
\label{sensitive2backbone}
\end{center}
\end{table}
1. \textbf{Sensitive to dimension}. The dimension of the last layer of the encoder (the dimension of representation) may influence the extracted EEG features contained in the representation and the task performance \cite{r50}. 2. \textbf{Sensitive to batch size}. The batch is another important parameter that influences training efficiency. 3. \textbf{Sensitive to the number of augmentation samples}. The inner-view and cross-view contrastive learning methods relied on the different augmented samples in the framework, and whether the number of the augmented samples will affect the 
performance of the model is an important content in the ablation experiment. 4. \textbf{Sensitive to backbone networks}. 
To verify the generalization of the self-supervised method proposed in this paper, the ablation experiment changed the backbone networks in the base encoders to show the stability of the framework. For the sensitivity to dimension, the dimension of 32, 64, 128, 256, 512, 1024, and 2048 were tested; For the sensitivity to batch size, the batch size of 16, 32, 64, 128, 256, 512, 1024 was tested; For the sensitivity to the number of augmentation samples, the number of 3-10 were tested. Besides, the different augmentation methods were tested: only frequency shuffle, only spatial shuffle, mix shuffle, only mask, and mixed methods. For the sensitive to backbone networks, this paper replaces the backbone network for the topology view from GCN to GCN fitted by Chebyshev inequality \cite{Chebnet}, GAT \cite{GAT}; and replaces the backbone network for the scalp view from CNN to the multi-layer fully connected neural networks.
\begin{table}[t]
\caption{The component ablation experiments on different datasets.}
\begin{center}
\label{table 3}
\begin{tabular}{ c c c c }
\hline
\textbf{Model} & \textbf{SEED} & \textbf{MMI} &\textbf{CHB-MIT} \\
\hline
Model-NP & 84.31$\%$ & 89.66$\%$ & 95.83$\%$ \\ 
Model-IN(T) & 91.25$\%$ & 91.34$\%$ & 97.99$\%$ \\
Model-IN(S) & 90.03$\%$ & 91.22$\%$ &97 84$\%$ \\
Model-IN(TS) & 91.07$\%$ & 92.37$\%$ & 98.17$\%$ \\
Model-CO  & 92.15$\%$ & 92.90$\%$ & 96.93$\%$ \\
Model-P & 89.74$\%$ & 90.82$\%$ & 97.35$\%$  \\
Model-S & 90.10$\%$ & 90.55$\%$ & 97.21$\%$ \\
\hline
\textbf{Model-normal} & \textbf{93.58$\%$} & \textbf{94.30$\%$} &\textbf{98.94$\%$}  \\
\hline
\end{tabular}
\label{abla_result}
\end{center}
\end{table}
Figure \ref{RQ4} and Table \ref{sensitive2backbone} show the result of the parameter ablation experiments. In Figure \ref{RQ4}(a), the result shows that increasing the vector dimension of representation can improve the neural knowledge expression ability to a certain extent, but if the dimension becomes too high, the task performance drops significantly. Dimensions 64 and 128 are most suitable for extracting neural features for different tasks. For the batch size, too small and big batch sizes are not helpful to the model, 128 is the best for the three tasks. Besides, too many augmentation samples in the framework may influence the efficiency of contrastive learning: Although adding augmentation samples will introduce more negative pairs in the cross-view contrastive, it also introduces too many positive pairs in the feature space of inner-view contrastive making it more difficult to mine invariant features from those abundant positive pairs. The imbalance of cross-view and inner-view contrastive loss poses challenges for the self-supervised training of models. From Figure \ref{RQ4}(d) and Table \ref{sensitive2backbone}, it can be concluded that the proposed framework is not sensitive to the backbone network and the augmentation method, different backbone networks and different combinations of augmentation methods achieved great results, which indicate the superior generalization of the proposed cross-view contrastive framework.

\section{Conclusion}
In this paper, we focus on the problem of the labeling issue that occurred in the EEG field and aim to extract the general implicit neural knowledge which can help solve different EEG-based tasks. Innovatively, this paper proposed a knowledge-driven cross-view contrastive learning method for EEG neurophysiological representation. The proposed framework is inspired by the multi-view learning and the contrastive learning theory, which constructs the hybrid contrastive learning strategies inner and cross two different views of EEG signal to self-supervised train the model and extract neural knowledge and improve the generalization of representation to adopt to different EEG-based downstream tasks. The experimental results on three different EEG-based tasks verified the superior performance of the proposed method: 1. The proposed method outperformed SOTA models on different tasks, indicating better task-generalization and neural representation ability. 2. It worked well on the few-label scenarios and can learn efficient features and knowledge through the non-label self-supervised training. 3. It is not sensitive to the backbone networks, which can combine different backbone models for different EEG signals as a general framework for EEG feature extraction.  In the future, more EEG neural knowledge can be integrated to model and analyze the EEG signals, constructing an interpretable general EEG representation framework for all EEG tasks. Furthermore, the structure can also be applied to other physiological computing to build a unified physiological signal representation architecture.


\begin{IEEEbiography}[{\includegraphics[width=1in,height=1.25in,clip,keepaspectratio]{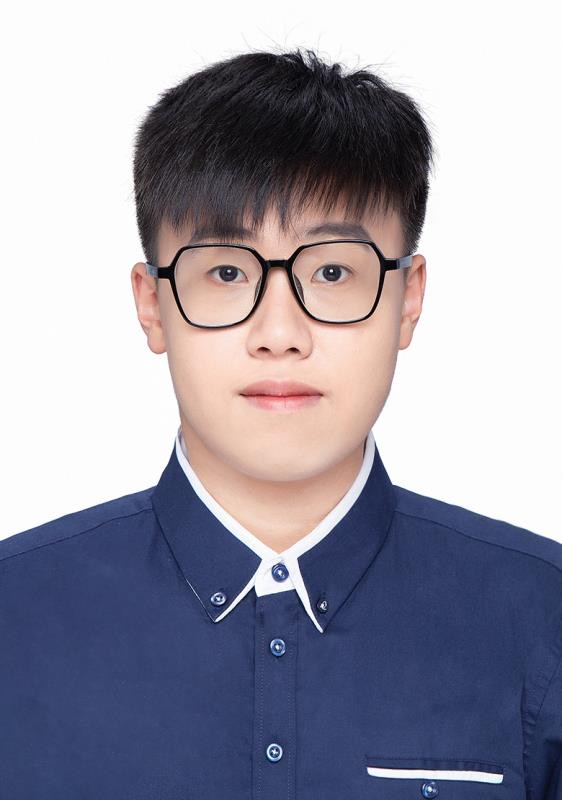}}]{Weining Weng} 
received the B.S. degree in Computer Science from Zhejiang University of Technology, Hangzhou, China, in 2021, and is now studying in University of Chinese Academy of Sciences for Ph.D degree. He is currently a Ph.D candidate of the Research Center for Ubiquitous Computing Systems (CUbiCS) at the Institute of Computing Technology (ICT), Chinese Academy of Sciences (CAS). His recent research mainly focuses on Self-supervised Learning, Perception Computing, and Intelligent Sensing.
\end{IEEEbiography}
%\begin{IEEEbiography}[{\includegraphics[width=1in,height=1.25in,clip,keepaspectratio]{yyhuang.png}
%}]{Yingying Huang} received the B.S. degree in Computer Science from Zhejiang University of Technology, Hangzhou, China, in 2021, and is now studying in Zhejiang University for Master's degree. She is currently a master candidate of  International Design Institute at Zhejiang University (ZJU). Her recent research mainly focuses on Human-computer Interaction .
%\end{IEEEbiography}
\begin{IEEEbiography}
[{\includegraphics[width=1in,height=1.25in,clip,keepaspectratio]{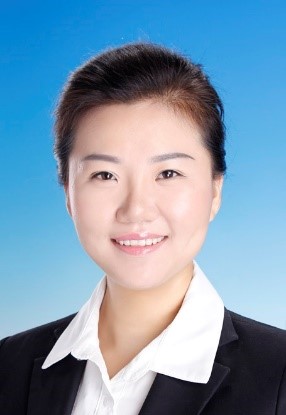}}]{Yang Gu} received the B.S. in computer science from Beijing University of Posts and Telecommunications, China in 2010, and Ph.D. degree in computer science from the Institute of Computing Technology (ICT), Chinese Academy of Sciences (CAS), Beijing, China, in 2016. She is currently an associate professor in the Research Center for Ubiquitous Computing Systems at ICT, CAS. Her research interests include Intelligent Sensing and Digital Health.
\end{IEEEbiography}

\begin{IEEEbiography}[{\includegraphics[width=1in,height=1.25in,clip,keepaspectratio]{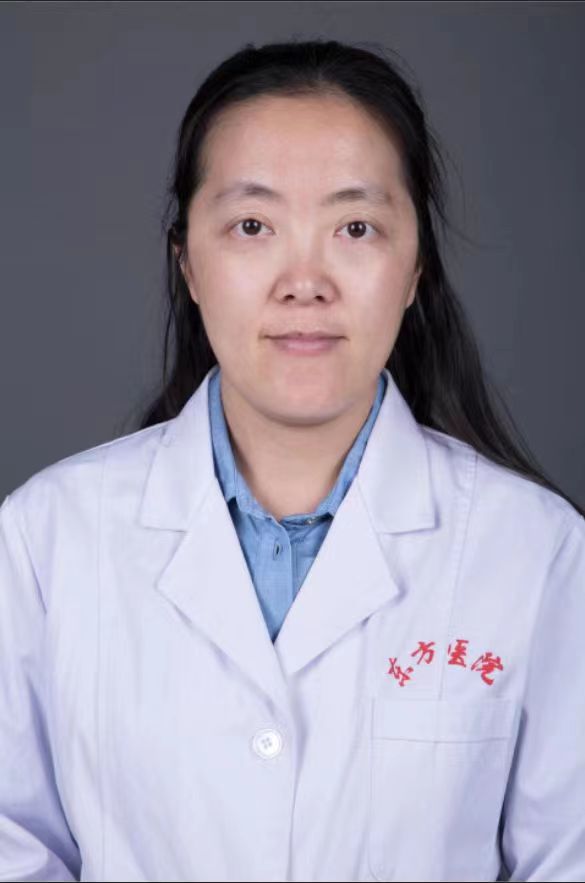}}]{Qihui Zhang} received the Ph.D degree from Beijing University of Chinese Medicine, Beijing, China. In 2006. She is a neurologist working at the Dongfang Hospital, Beijing University of Chinese Medicine. Beijing, China, since 2006. She worked her medical research in cerebrovascular disease research as a Post-doctoral visiting scholar under the supervision of Dr. Jeffrey Saver, at the UCLA department of neurology, UCLA Comprehensive Stroke Center, USA, from January 2012 to February 2013.  She continued her neurology research at Beijing Tiantan Hospital, Beijing, China, as a postdoctor, from 2017 to 2021. She worked at the Department of neurology, foothills medical centre, University of Calgary, Canada, as a Visiting Professor/Researcher, from 2020 to 2022. Her research interests are in acute stroke treatment, stroke prevention, neuroimaging. 
\end{IEEEbiography}
\begin{IEEEbiography}[{\includegraphics[width=1in,height=1.25in,clip,keepaspectratio]{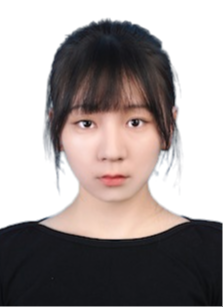}
}]{Yingying Huang} received the B.S. degree in Computer Science from Zhejiang University of Technology, Hangzhou, China, in 2021, and is now studying in Zhejiang University for Master's degree. She is currently a master candidate of  International Design Institute at Zhejiang University (ZJU). Her recent research mainly focuses on Human-computer Interaction .
\end{IEEEbiography}
\begin{IEEEbiography}[{\includegraphics[width=1in,height=1.25in,clip,keepaspectratio]{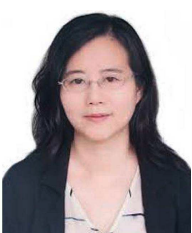}}]{Chunyan Miao} is the Director of the Joint NTUUBC Research Centre of Excellence in Active Living for the Elderly (LILY), Nanyang Technological University (NTU), Singapore. She is the Chair of the School of Computer Science and Engineering, NTU. She is the Editor-in-Chief of the International Journal of Information Technology published by the Singapore Computer Society.
\end{IEEEbiography}
\begin{IEEEbiography}
[{\includegraphics[width=1in,height=1.25in,clip,keepaspectratio]{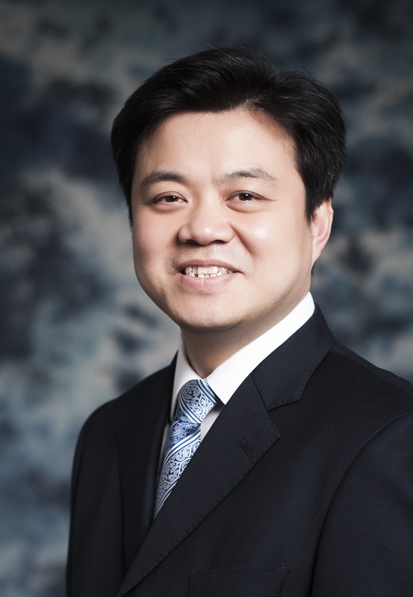}}]{Yiqiang Chen}
received the B.S. and M.S. degrees in computer science from Xiangtan University, Xiangtan, China, in 1996 and 1999, respectively, and the Ph.D. degree in computer science from the Institute of Computing Technology, Chinese Academy of Sciences, Beijing, China, in 2003. In 2004, he was a Visiting Scholar Researcher with the Department of Computer
Science, Hong Kong University of Science and Technology (HKUST), Hong Kong. He is currently a professor and the director of the Research Center for Ubiquitous Computing Systems at the Institute of Computing Technology (ICT), Chinese Academy of Sciences (CAS). His research interests include artificial intelligence, pervasive computing, and human-computer interaction.
\end{IEEEbiography}


\begin{thebibliography}{60}

\bibitem{r1}
D.~P. Subha, P.~K. Joseph, R.~Acharya~U, and C.~M. Lim, ``Eeg signal analysis:
  a survey,'' {\em Journal of medical systems}, vol.~34, pp.~195--212, 2010.

\bibitem{r2}
A.~F. Jackson and D.~J. Bolger, ``The neurophysiological bases of eeg and eeg
  measurement: A review for the rest of us,'' {\em Psychophysiology}, vol.~51,
  no.~11, pp.~1061--1071, 2014.

\bibitem{r3}
A.~Markovic, M.~Kaess, and L.~Tarokh, ``Gender differences in adolescent sleep
  neurophysiology: a high-density sleep eeg study,'' {\em Scientific reports},
  vol.~10, no.~1, pp.~1--13, 2020.

\bibitem{r4}
T.~K.~K. Ho and N.~Armanfard, ``Self-supervised learning for anomalous channel
  detection in eeg graphs: Application to seizure analysis,'' {\em arXiv
  preprint arXiv:2208.07448}, 2022.

\bibitem{r5}
S.~Lee, Y.~Yu, S.~Back, H.~Seo, and K.~Lee, ``Sleepyco: Automatic sleep scoring
  with feature pyramid and contrastive learning,'' {\em arXiv preprint
  arXiv:2209.09452}, 2022.

\bibitem{r6}
N.~S. E.~M. Noor and H.~Ibrahim, ``Machine learning algorithms and quantitative
  electroencephalography predictors for outcome prediction in traumatic brain
  injury: A systematic review,'' {\em IEEE Access}, vol.~8, pp.~102075--102092,
  2020.

%\bibitem{r7}
%J.~N. Ianof and R.~Anghinah, ``Traumatic brain injury: An eeg point of view,''
%  {\em Dementia \& neuropsychologia}, vol.~11, pp.~3--5, 2017.

\bibitem{r8}
A.~Dietrich and R.~Kanso, ``A review of eeg, erp, and neuroimaging studies of
  creativity and insight.,'' {\em Psychological bulletin}, vol.~136, no.~5,
  p.~822, 2010.

\bibitem{r9}
X.-W. Wang, D.~Nie, and B.-L. Lu, ``Emotional state classification from eeg
  data using machine learning approach,'' {\em Neurocomputing}, vol.~129,
  pp.~94--106, 2014.

%\bibitem{r10}
%P.~C. Petrantonakis and L.~J. Hadjileontiadis, ``Emotion recognition from eeg
%  using higher order crossings,'' {\em IEEE Transactions on information
%  Technology in Biomedicine}, vol.~14, no.~2, pp.~186--197, 2009.

%\bibitem{r11}
%U.~R. Acharya, S.~V. Sree, G.~Swapna, R.~J. Martis, and J.~S. Suri, ``Automated
%  eeg analysis of epilepsy: a review,'' {\em Knowledge-Based Systems}, vol.~45,
%  pp.~147--165, 2013.

\bibitem{r12}
A.~Craik, Y.~He, and J.~L. Contreras-Vidal, ``Deep learning for
  electroencephalogram (eeg) classification tasks: a review,'' {\em Journal of
  neural engineering}, vol.~16, no.~3, p.~031001, 2019.

%\bibitem{r13}
%X.~Zhang and D.~Wu, ``On the vulnerability of cnn classifiers in eeg-based
%  bcis,'' {\em IEEE transactions on neural systems and rehabilitation
%  engineering}, vol.~27, no.~5, pp.~814--825, 2019.

%\bibitem{r14}
%S.~Alhagry, A.~A. Fahmy, and R.~A. El-Khoribi, ``Emotion recognition based on
%  eeg using lstm recurrent neural network,'' {\em International Journal of
%  Advanced Computer Science and Applications}, vol.~8, no.~10, 2017.

\bibitem{r15}
Y.~Song, X.~Jia, L.~Yang, and L.~Xie, ``Transformer-based spatial-temporal
  feature learning for eeg decoding,'' {\em arXiv preprint arXiv:2106.11170},
  2021.

\bibitem{r16}
R.~Zhou, Z.~Zhang, X.~Yang, H.~Fu, L.~Zhang, L.~Li, G.~Huang, Y.~Dong, F.~Li,
  and Z.~Liang, ``A novel transfer learning framework with prototypical
  representation based pairwise learning for cross-subject cross-session
  eeg-based emotion recognition,'' {\em arXiv preprint arXiv:2202.06509}, 2022.

\bibitem{r17}
M.~H. Rafiei, L.~V. Gauthier, H.~Adeli, and D.~Takabi, ``Self-supervised
  learning for electroencephalography,'' {\em IEEE Transactions on Neural
  Networks and Learning Systems}, 2022.

\bibitem{r18}
H.-Y.~S. Chien, H.~Goh, C.~M. Sandino, and J.~Y. Cheng, ``Maeeg: Masked
  auto-encoder for eeg representation learning,'' {\em arXiv preprint
  arXiv:2211.02625}, 2022.

\bibitem{r20}
J.~Han, X.~Gu, and B.~Lo, ``Semi-supervised contrastive learning for
  generalizable motor imagery eeg classification,'' in {\em 2021 IEEE 17th
  International Conference on Wearable and Implantable Body Sensor Networks
  (BSN)}, pp.~1--4, IEEE, 2021.

\bibitem{r19}
Y.~Li, J.~Chen, F.~Li, B.~Fu, H.~Wu, Y.~Ji, Y.~Zhou, Y.~Niu, G.~Shi, and
  W.~Zheng, ``Gmss: Graph-based multi-task self-supervised learning for eeg
  emotion recognition,'' {\em IEEE Transactions on Affective Computing}, 2022.

\bibitem{r21}
M.~Yang, Y.~Li, Z.~Huang, Z.~Liu, P.~Hu, and X.~Peng, ``Partially view-aligned
  representation learning with noise-robust contrastive loss,'' in {\em
  Proceedings of the IEEE/CVF conference on computer vision and pattern
  recognition}, pp.~1134--1143, 2021.

\bibitem{r22}
Q.~Wen, Z.~Ouyang, C.~Zhang, Y.~Qian, Y.~Ye, and C.~Zhang, ``Graph contrastive
  learning with cross-view reconstruction,'' in {\em NeurIPS 2022 Workshop: New
  Frontiers in Graph Learning}.

%\bibitem{r23}
%K.~Hassani and A.~H. Khasahmadi, ``Contrastive multi-view representation
%  learning on graphs,'' in {\em International conference on machine learning},
%  pp.~4116--4126, PMLR, 2020.

\bibitem{r24}
D.~Zou, W.~Wei, X.-L. Mao, Z.~Wang, M.~Qiu, F.~Zhu, and X.~Cao, ``Multi-level
  cross-view contrastive learning for knowledge-aware recommender system,'' in
  {\em Proceedings of the 45th International ACM SIGIR Conference on Research
  and Development in Information Retrieval}, pp.~1358--1368, 2022.

%\bibitem{r25}
%Y.~Ma, Y.~He, A.~Zhang, X.~Wang, and T.-S. Chua, ``Crosscbr: Cross-view
%  contrastive learning for bundle recommendation,'' {\em arXiv preprint
%  arXiv:2206.00242}, 2022.

\bibitem{r26}
S.~K. Khare and V.~Bajaj, ``Time--frequency representation and convolutional
  neural network-based emotion recognition,'' {\em IEEE transactions on neural
  networks and learning systems}, vol.~32, no.~7, pp.~2901--2909, 2020.

\bibitem{r27}
W.~Tao, C.~Li, R.~Song, J.~Cheng, Y.~Liu, F.~Wan, and X.~Chen, ``Eeg-based
  emotion recognition via channel-wise attention and self attention,'' {\em
  IEEE Transactions on Affective Computing}, 2020.

%\bibitem{r28}
%M.~Jin, E.~Zhu, C.~Du, H.~He, and J.~Li, ``Pgcn: Pyramidal graph convolutional
%  network for eeg emotion recognition,'' {\em arXiv preprint arXiv:2302.02520},
%  2023.

\bibitem{r29}
H.~Altaheri, G.~Muhammad, M.~Alsulaiman, S.~U. Amin, G.~A. Altuwaijri,
  W.~Abdul, M.~A. Bencherif, and M.~Faisal, ``Deep learning techniques for
  classification of electroencephalogram (eeg) motor imagery (mi) signals: A
  review,'' {\em Neural Computing and Applications}, pp.~1--42, 2021.

\bibitem{r30}
S.~U. Amin, M.~Alsulaiman, G.~Muhammad, M.~A. Bencherif, and M.~S. Hossain,
  ``Multilevel weighted feature fusion using convolutional neural networks for
  eeg motor imagery classification,'' {\em Ieee Access}, vol.~7,
  pp.~18940--18950, 2019.

%\bibitem{r31}
%M.~Dai, D.~Zheng, R.~Na, S.~Wang, and S.~Zhang, ``Eeg classification of motor
%  imagery using a novel deep learning framework,'' {\em Sensors}, vol.~19,
%  no.~3, p.~551, 2019.

\bibitem{r32}
Y.~Hou, S.~Jia, X.~Lun, Z.~Hao, Y.~Shi, Y.~Li, R.~Zeng, and J.~Lv, ``Gcns-net:
  a graph convolutional neural network approach for decoding time-resolved eeg
  motor imagery signals,'' {\em IEEE Transactions on Neural Networks and
  Learning Systems}, 2022.

%\bibitem{r33}
%X.~Zheng and W.~Chen, ``An attention-based bi-lstm method for visual %object
%  classification via eeg,'' {\em Biomedical Signal Processing and Control},
%  vol.~63, p.~102174, 2021.

%\bibitem{r34}
%X.~Zheng, W.~Chen, Y.~You, Y.~Jiang, M.~Li, and T.~Zhang, ``Ensemble deep
%  learning for automated visual classification using eeg signals,'' {\em
%  Pattern Recognition}, vol.~102, p.~107147, 2020.

\bibitem{r35}
S.~Madhavan, R.~K. Tripathy, and R.~B. Pachori, ``Time-frequency domain deep
  convolutional neural network for the classification of focal and non-focal
  eeg signals,'' {\em IEEE Sensors Journal}, vol.~20, no.~6, pp.~3078--3086,
  2019.

\bibitem{r36}
J.~Sun, R.~Cao, M.~Zhou, W.~Hussain, B.~Wang, J.~Xue, and J.~Xiang, ``A hybrid
  deep neural network for classification of schizophrenia using eeg data,''
  {\em Scientific Reports}, vol.~11, no.~1, pp.~1--16, 2021.

\bibitem{r37}
H.~Banville, O.~Chehab, A.~Hyv{\"a}rinen, D.-A. Engemann, and A.~Gramfort,
  ``Uncovering the structure of clinical eeg signals with self-supervised
  learning,'' {\em Journal of Neural Engineering}, vol.~18, no.~4, p.~046020,
  2021.

%\bibitem{r38}
%H.~Banville, I.~Albuquerque, A.~Hyv{\"a}rinen, G.~Moffat, D.-A. Engemann, and
%  A.~Gramfort, ``Self-supervised representation learning from
%  electroencephalography signals,'' in {\em 2019 IEEE 29th International
%  Workshop on Machine Learning for Signal Processing (MLSP)}, pp.~1--6, IEEE,
%  2019.

\bibitem{r39}
J.~Xu, Y.~Zheng, Y.~Mao, R.~Wang, and W.-S. Zheng, ``Anomaly detection on
  electroencephalography with self-supervised learning,'' in {\em 2020 IEEE
  International Conference on Bioinformatics and Biomedicine (BIBM)},
  pp.~363--368, IEEE, 2020.

%\bibitem{r40}
%Z.~Zhang, S.-h. Zhong, and Y.~Liu, ``Ganser: A self-supervised data
%  augmentation framework for eeg-based emotion recognition,'' {\em IEEE
%  Transactions on Affective Computing}, 2022.

\bibitem{r41}
D.~Kostas, S.~Aroca-Ouellette, and F.~Rudzicz, ``Bendr: using transformers and
  a contrastive self-supervised learning task to learn from massive amounts of
  eeg data,'' {\em Frontiers in Human Neuroscience}, vol.~15, p.~653659, 2021.

\bibitem{r42}
R.~Li, Y.~Wang, W.-L. Zheng, and B.-L. Lu, ``A multi-view
  spectral-spatial-temporal masked autoencoder for decoding emotions with
  self-supervised learning,'' in {\em Proceedings of the 30th ACM International
  Conference on Multimedia}, pp.~6--14, 2022.

\bibitem{r43}
M.~N. Mohsenvand, M.~R. Izadi, and P.~Maes, ``Contrastive representation
  learning for electroencephalogram classification,'' in {\em Machine Learning
  for Health}, pp.~238--253, PMLR, 2020.

\bibitem{r44}
H.~Kan, J.~Yu, J.~Huang, Z.~Liu, and H.~Zhou, ``Self-supervised group meiosis
  contrastive learning for eeg-based emotion recognition,'' {\em arXiv preprint
  arXiv:2208.00877}, 2022.

\bibitem{r45}
X.~Shen, X.~Liu, X.~Hu, D.~Zhang, and S.~Song, ``Contrastive learning of
  subject-invariant eeg representations for cross-subject emotion
  recognition,'' {\em IEEE Transactions on Affective Computing}, 2022.

\bibitem{r46}
F.~Deligianni, M.~Centeno, D.~W. Carmichael, and J.~D. Clayden, ``Relating
  resting-state fmri and eeg whole-brain connectomes across frequency bands,''
  {\em Frontiers in neuroscience}, vol.~8, p.~258, 2014.

\bibitem{r47}
R.-N. Duan, J.-Y. Zhu, and B.-L. Lu, ``Differential entropy feature for
  eeg-based emotion classification,'' in {\em 2013 6th International IEEE/EMBS
  Conference on Neural Engineering (NER)}, pp.~81--84, IEEE, 2013.

\bibitem{r48}
K.~He, X.~Chen, S.~Xie, Y.~Li, P.~Doll{\'a}r, and R.~Girshick, ``Masked
  autoencoders are scalable vision learners,'' in {\em Proceedings of the
  IEEE/CVF Conference on Computer Vision and Pattern Recognition},
  pp.~16000--16009, 2022.

\bibitem{r49}
M.~Noroozi and P.~Favaro, ``Unsupervised learning of visual representations by
  solving jigsaw puzzles,'' in {\em Computer Vision--ECCV 2016: 14th European
  Conference, Amsterdam, The Netherlands, October 11-14, 2016, Proceedings,
  Part VI}, pp.~69--84, Springer, 2016.

\bibitem{r50}
J.~Zbontar, L.~Jing, I.~Misra, Y.~LeCun, and S.~Deny, ``Barlow twins:
  Self-supervised learning via redundancy reduction,'' in {\em International
  Conference on Machine Learning}, pp.~12310--12320, PMLR, 2021.

\bibitem{r51}
T.~Chen, S.~Kornblith, M.~Norouzi, and G.~Hinton, ``A simple framework for
  contrastive learning of visual representations,'' in {\em International
  conference on machine learning}, pp.~1597--1607, PMLR, 2020.

\bibitem{r52}
A.~Kendall, Y.~Gal, and R.~Cipolla, ``Multi-task learning using uncertainty to
  weigh losses for scene geometry and semantics,'' in {\em Proceedings of the
  IEEE conference on computer vision and pattern recognition}, pp.~7482--7491,
  2018.

\bibitem{r53}
W.-L. Zheng and B.-L. Lu, ``Investigating critical frequency bands and channels
  for eeg-based emotion recognition with deep neural networks,'' {\em IEEE
  Transactions on autonomous mental development}, vol.~7, no.~3, pp.~162--175,
  2015.

\bibitem{r54}
G.~Schalk, D.~J. McFarland, T.~Hinterberger, N.~Birbaumer, and J.~R. Wolpaw,
  ``Bci2000: a general-purpose brain-computer interface (bci) system,'' {\em
  IEEE Transactions on biomedical engineering}, vol.~51, no.~6, pp.~1034--1043,
  2004.

\bibitem{r55}
A.~H. Shoeb, {\em Application of machine learning to epileptic seizure onset
  detection and treatment}.
\newblock PhD thesis, Massachusetts Institute of Technology, 2009.

\bibitem{r56}
J.-B. Grill, F.~Strub, F.~Altch{\'e}, C.~Tallec, P.~Richemond, E.~Buchatskaya,
  C.~Doersch, B.~Avila~Pires, Z.~Guo, M.~Gheshlaghi~Azar, {\em et~al.},
  ``Bootstrap your own latent-a new approach to self-supervised learning,''
  {\em Advances in neural information processing systems}, vol.~33,
  pp.~21271--21284, 2020.

\bibitem{r57}
K.~He, H.~Fan, Y.~Wu, S.~Xie, and R.~Girshick, ``Momentum contrast for
  unsupervised visual representation learning,'' in {\em Proceedings of the
  IEEE/CVF conference on computer vision and pattern recognition},
  pp.~9729--9738, 2020.

\bibitem{r58}
X.~Chen and K.~He, ``Exploring simple siamese representation learning,'' in
  {\em Proceedings of the IEEE/CVF conference on computer vision and pattern
  recognition}, pp.~15750--15758, 2021.

\bibitem{r59}
Y.~Ou, S.~Sun, H.~Gan, R.~Zhou, and Z.~Yang, ``An improved self-supervised
  learning for eeg classification,'' {\em Mathematical Biosciences and
  Engineering}, vol.~19, no.~7, pp.~6907--6922, 2022.

\bibitem{r60}
F.~Shen, G.~Dai, G.~Lin, J.~Zhang, W.~Kong, and H.~Zeng, ``Eeg-based emotion
  recognition using 4d convolutional recurrent neural network,'' {\em Cognitive
  Neurodynamics}, vol.~14, pp.~815--828, 2020.

\bibitem{Chebnet}
M.~Defferrard, X.~Bresson, and P.~Vandergheynst, ``Convolutional neural
  networks on graphs with fast localized spectral filtering,'' {\em Advances in
  neural information processing systems}, vol.~29, 2016.

\bibitem{GAT}
P.~Veli{\v{c}}kovi{\'c}, G.~Cucurull, A.~Casanova, A.~Romero, P.~Lio, and
  Y.~Bengio, ``Graph attention networks,'' {\em arXiv preprint
  arXiv:1710.10903}, 2017.

\end{thebibliography}
\end{document}